\documentclass[aps,pra,10pt,showpacs,tightenlines,longbibliography,notitlepage,superscriptaddress,twocolumn,floatfix,eqsecnum]{revtex4-1}

\usepackage{graphicx,amssymb,amstext}
\usepackage{amsmath}
\usepackage{hyperref}
\usepackage{epsfig}
\usepackage{verbatim}

\newtheorem{theorem}{Theorem}[section]

\begin{document}


\title{Robust Guaranteed-Cost Adaptive Quantum Phase Estimation}

\author{Shibdas Roy}
\email{roy\_shibdas@yahoo.co.in}
\affiliation{Department of Physics, University of Warwick, United Kingdom}
\affiliation{Department of Electrical and Computer Engineering, National University of Singapore, Singapore}
\author{Dominic W. Berry}
\email{dmwberry@gmail.com}
\affiliation{Department of Physics and Astronomy, Macquarie University, Sydney, Australia}
\author{Ian R. Petersen}
\email{i.r.petersen@gmail.com}
\affiliation{Research School of Engineering, Australian National University, Canberra, Australia}
\author{Elanor H. Huntington}
\email{elanor.huntington@anu.edu.au}
\affiliation{Research School of Engineering, Australian National University, Canberra, Australia}
\affiliation{Centre for Quantum Computation and Communication Technology, Australian Research Council}



\begin{abstract}
Quantum parameter estimation plays a key role in many fields like quantum computation, communication and metrology. Optimal estimation allows one to achieve the most precise parameter estimates, but requires accurate knowledge of the model. Any inevitable uncertainty in the model parameters may heavily degrade the quality of the estimate. It is therefore desired to make the estimation process robust to such uncertainties. Robust estimation was previously studied for a varying phase, where the goal was to estimate the phase at some time in the past, using the measurement results from both before and after that time within a fixed time interval upto current time. Here, we consider a robust guaranteed-cost filter yielding robust estimates of a varying phase in real time, where the current phase is estimated using only past measurements. Our filter minimizes the largest (worst-case) variance in the allowable range of the uncertain model parameter(s) and this determines its guaranteed cost. It outperforms in the worst case the optimal Kalman filter designed for the model with no uncertainty, that corresponds to the center of the possible range of the uncertain parameter(s). Moreover, unlike the Kalman filter, our filter in the worst case always performs better than the best achievable variance for heterodyne measurements, that we consider as the tolerable threshold for our system. Furthermore, we consider effective quantum efficiency and effective noise power, and show that our filter provides the best results by these measures in the worst case.
\end{abstract}


\maketitle

\section{Introduction}
Quantum parameter estimation plays a central role in many fields, such as quantum computing \cite{HWA}, communication \cite{SPK,IWY,CHD} and metrology \cite{GLM1}. It involves estimating a classical variable, such as an optical phase shift, of a quantum system \cite{WM}. This is required since measurements are corrupted with unavoidable noises in practice and it is important to estimate as accurately as possible the desired variable from such noisy measurements. Also, often the variable of interest is not directly accessible to measurement. As an example, the detection of an optical beam always results in the measurement of the amplitude of the beam. It is, however, an important task to estimate the phase of an optical beam from the available measurement results for the amplitude of the field. In communications, for example, it is more reliable to encode information into the phase rather than the amplitude or intensity of the optical field \cite{EHH}. Moreover, adaptive measurements (i.e.~involving feedback) allow for demonstrably better estimation of phase than non-adaptive estimation. Quantum phase estimation has therefore been extensively studied in this context \cite{HMW,BW1,BW2,TSL,PWL,WK1,WK2,MA}.

Optimal estimation strategies have played significant roles in improving the achievable precision in quantum phase estimation \cite{TW,YNW,KI}. For example, adaptive estimation with a Kalman filter has provided mean-square errors less than the standard quantum limit (SQL) \cite{TW,RPH2}. The SQL is the minimum phase estimation error that can be obtained with a coherent beam using a perfect heterodyne technique and sets an important benchmark for the quality of a measurement. However, optimal estimation schemes provide optimal performance only if the parameters underlying the system are accurately known. Any uncertainty in our knowledge of these parameters can heavily degrade the estimation performance under certain circumstances. In the presence of large uncertainty, the worst-case mean-square errors can be higher than the acceptable thresholds in practical engineering systems. Such a threshold often determines the point beyond which a system has a risk of breaking down or becoming unusable. It is therefore important to make the estimation process robust to such uncertainties \cite{LXP,ZDG}. A robust filter is one that provides acceptable performance for the full possible range of the uncertain system parameter(s), and the worst case is the situation where the parameters result in the largest mean-square error.

For quantum estimation problems that can be approximated as having linearized dynamics, many of the principles of robust classical estimation can be applied. There is a rich set of classical estimation strategies in modern control theory for achieving robustness for systems with explicitly introduced uncertainties in a systematic state-space setting \cite{LXP,ZDG}. Such classical robust estimation principles as applied to quantum estimation problems are little studied. Some of the authors have earlier studied in Ref.~\cite{RPH1} the problem of robustly estimating a continuously-varying phase, modelled as a stochastic noisy process, of a squeezed state of light \cite{RPH6,RPH5}. In particular, a robust fixed-interval smoother was constructed, that provides guaranteed worst-case performance when compared to an optimal smoother in adaptive quantum optical phase estimation. A smoother involves deducing an estimate based on both past and future measurements with respect to the time at which the phase is to be estimated \cite{MT1,MT2}. This non-causal estimator is of relevance to applications such as gravitational wave detection \cite{GMM,LIGO}, where obtaining more precise estimates is of greater interest than real-time estimates.

However, for applications such as quantum computing and communication, it is instead required to obtain real-time estimates as precisely as possible. In such cases, it is filtering, involving only past measurements, that can be used and should be made to yield as accurate real-time estimates as possible. In this paper, we consider a robust guaranteed-cost filter \cite{PM} in an adaptive quantum phase estimation process of the form considered in Ref.~\cite{TW}, using a coherent state of light. Such a robust filter helps in achieving real-time guaranteed worst-case performance, in relation to the SQL taken as the tolerable threshold, over a Kalman filter. The phase to be estimated is modelled as an Ornstein-Uhlenbeck (OU) process (Eq.~(\ref{eq:ou_process})) as in Refs.~\cite{TW,YNW,RPH1}.

The robust smoother illustrated in Ref.~\cite{RPH1} was shown to consist of a forward robust filter and a backward robust filter, the estimates of which were combined to yield the desired smoothed estimate. However, the approach of Ref.~\cite{RPH1} considered an energy bounded description of noise \cite{MSP}. By contrast, we here use the approach of Ref.~\cite{PM} that considered a white noise description, which is the actual noise involved here (Eqs.~(\ref{eq:exact_sys_model}), (\ref{eq:model_uncorrelated})). In addition, such a robust filter (Eq.~(\ref{eq:robust_filter})), as its name suggests, comes with a guaranteed cost (Eq.~(\ref{eq:robust_worst_error})), that quantifies an upper bound on the mean-square error of estimates obtained from the filter for the uncertain system (Eq.~(\ref{eq:robust_error_worst})). In other words, the worst-case mean-square error of the robust filter is guaranteed to be bounded by the aforementioned cost.

The robust guaranteed-cost filter considered here has previously been applied by three of the authors to a coherent state of light to yield some preliminary results \cite{RPH2,RPH3}. This paper builds on Ref.~\cite{RPH2} and presents more in-depth results, insights and analysis relevant to the physical problem at hand. We demonstrate that with uncertainty in the system (Eq.~(\ref{eq:unc_sys_model})), the Kalman filter performance (Eq.~(\ref{eq:kalman_error_worst})) gets increasingly worse with respect to the SQL (Eq.~(\ref{eq:ou_sql})), whereas the robust filter (Eq.~(\ref{eq:robust_error_worst})) consistently beats the SQL in the worst-case scenario. We explicitly illustrate that the robust filter is guaranteed to beat the Kalman filter as well as the SQL in the worst-case situation.

Moreover, we define an effective quantum efficiency (Eq.~(\ref{eq:eff_quantum_efficiency})) for our filters, such that the best estimation scheme yields the same variance as the suboptimal estimation scheme with no loss \cite{HBL}. We illustrate that our robust filter always exhibits the maximum achievable effective quantum efficiency in the worst-error case, whereas the Kalman filter suffers heavily in this regard. Furthermore, we define an effective noise power (Eqs.~(\ref{eq:kappa_eff}), (\ref{eq:kappa_n})) for our filters, such that the mean-square error for the optimal filter with added noise is the same as that for a suboptimal filter. This is inspired by a similar result for a constant phase \cite{MJWH1,MJWH2}. We see that our robust filter always admits the minimum possible effective noise power in the worst-error case, but the Kalman filter suffers from much higher effective noise owing to the uncertainty in the model.

The structure of the paper is as follows. Section \ref{sec:est_filters} provides an introduction to the theory of estimation using filters. Section \ref{sec:opt_filter} discusses the optimal adaptive estimation process using a Kalman filter for an exact system model. Section \ref{sec:rob_filter} deduces the robust filter for the uncertain system model. Section \ref{sec:lyapunov} shows how to calculate the filter mean-square errors for the uncertain system. Section \ref{sec:ou_comp_errors} concerns the comparison of the mean-square errors of the Kalman and robust filters for the uncertain system. Section \ref{sec:error_analysis} illustrates that the worst-case mean-square error of the robust filter always beats that of the Kalman filter as well as the SQL. Section \ref{sec:ou_eff_qeff} introduces a quantity called the effective quantum efficiency for the Kalman and robust filters and compares the filters in terms of this quantity. Section \ref{sec:ou_eff_np} defines another quantity called the effective noise power for the filters, that are then compared in terms of this quantity. Finally, Section \ref{sec:concl} ends the paper with concluding remarks.

\vspace*{-2mm}
\section{Estimation Using Filters}\label{sec:est_filters}
Here we look at the problem of estimating a set of time-varying parameters describing the state of a system, given noisy measurements (output) of observable quantities along with a model relating the observations to the underlying state, corrupted by noise (input). Such a dynamical system is conveniently modelled in a compact state-space form as a set of input, output and state variables related by first order differential equations, where the variables are expressed as vectors. If the system is linear and time-invariant, the differential and algebraic equations are written in matrix form as follows \cite{SR}:
\begin{align}
\dot{x}(t) &= Ax(t) + Bv(t),\label{eq:state_space1}\\
y(t) &= Cx(t) + Dw(t),\label{eq:state_space2}\\
z(t) &= Lx(t).\label{eq:state_space3}
\end{align}
Here $x$ is the state vector, $y$ is the output vector, $z$ is the vector of variables to be estimated, $v$ and $w$ are the inputs and are (possibly vector) white noise processes that may be correlated, $t$ is the time variable, and $A$, $B$, $C$, $D$ and $L$ are suitable matrices of compatible dimensions. In particular, $A$ is referred to as the state matrix.

A suitable estimator for the above system (\ref{eq:state_space1})-(\ref{eq:state_space3}) is then a filter of the form:
\begin{align}
\dot{\hat{x}}(t) &= F\hat{x}(t) + Ky(t),\label{eq:filter_form1}\\
\hat{z}(t) &= L\hat{x}(t)\label{eq:filter_form2},
\end{align}
where $\hat{x}$ is the estimate of the state $x$ and $\hat{z}$ is the estimate of the desired quantity $z$. The input to the filter is the output vector $y$ of the system, and the filter output is the estimate $\hat{z}$. In the Laplace domain, we get:
\begin{align}
(sI-F)\hat{x}(s) &= Ky(s),\\
\hat{z}(s) &= L\hat{x}(s),
\end{align}
where $s$ is the Laplace variable. The transfer function, relating the filter output to the filter input, is therefore,
\begin{equation}\label{eq:filter_tf}
G(s) = \frac{\hat{z}(s)}{y(s)} = L(sI-F)^{-1}K.
\end{equation}

The estimation error is $\varepsilon = z - \hat{z}$. Then, the optimal steady-state filter is the system (\ref{eq:filter_form1}), (\ref{eq:filter_form2}), (\ref{eq:filter_tf}) with $F$ and $K$ suitably chosen to minimize the cost,
\begin{equation}\label{eq:filter_cost}
J_c = \lim_{t \to \infty} \langle \varepsilon^T(t)\varepsilon(t) \rangle ,
\end{equation}
where $\langle\cdot\rangle$ denotes the expectation.

The optimal estimator that minimizes the above cost is given by the Kalman filter. It operates recursively on streams of noisy input data to produce a statistically optimal estimate of the underlying system state. The steady-state Kalman filter (\ref{eq:filter_form1}), (\ref{eq:filter_form2}) for the system (\ref{eq:state_space1})-(\ref{eq:state_space3}) is obtained from the solution to the algebraic Riccati equation \cite{RPH1,LXP,RGB}:
\begin{equation}\label{eq:kalman_filter_riccati}
AP+PA^T+BQB^T-PC^T(DRD^T)^{-1}CP=0,
\end{equation}
which is quadratic in $P$, the error-covariance matrix for the state. The positive semi-definite solution (if one exists) for $P$ is the stabilising solution that is desired. Here, we have assumed the following about the white noises $v$ and $w$ (in particular that they are uncorrelated):
\begin{align}
\langle v(t)v^T(\tau)\rangle &= Q\delta(t-\tau),\\
\langle w(t)w^T(\tau)\rangle &= R\delta(t-\tau),\\
\langle v(t)w^T(\tau)\rangle &= 0,
\end{align}
where $\delta(\cdot)$ is the Dirac delta function. Then, the Kalman filter cost (\ref{eq:filter_cost}) is:
\begin{equation}\label{eq:kalman_cost}
J_c = LPL^T,
\end{equation}
which is the error-covariance for the desired quantity, and the relevant matrices of the filter are:
\begin{align}
F &= A-KC,\label{eq:kalman_filter_coeffs1}\\
K &= PC^T(DRD^T)^{-1}.\label{eq:kalman_filter_coeffs2}
\end{align}
Here the matrix $K$ is the gain matrix of the Kalman filter, when $L$ is identity, and therefore, is commonly referred to as the Kalman gain.

Essentially, the Kalman filter recursively calculates a new prediction of the system's state upon averaging it with a new measurement, weighted suitably based on the estimated certainty: values with higher certainty are trusted more. The Kalman filter needs only the last ``best guess" instead of the entire history of a system's state to calculate a new one. The Kalman gain determines the relative certainty of the current state and the measurements, and therefore, the filter performance \cite{SR}.

\section{Optimal Filter}\label{sec:opt_filter}
The adaptive estimator used in Ref.~\cite{TW} involves a feedback filter and an offline smoother, that yields the final optimal estimate with a delay with respect to the estimation time. The feedback filter used in Ref.~\cite{TW} is suboptimal and would be optimal when using a Kalman filter \cite{RPH2}. Here, we shall consider the Kalman filter alone, the output of which is the desired optimal real-time estimate, that is also used to adapt the local oscillator phase through feedback in the adaptive estimation process. First, we define the process and measurement models for our system in a state-space setting \cite{RPH1}.

\subsection{System Model (Exact)}
An Ornstein-Uhlenbeck (OU) noise process modulates the phase $\phi(t)$, to be estimated, of the continuous coherent beam of light \cite{TW} and constitutes our process model:
\begin{equation}\label{eq:ou_process}
\dot{\phi}(t) = -\lambda\phi(t)+\sqrt{\kappa}v(t),
\end{equation}
where $\lambda > 0$ is the inverse correlation time of the phase, $\kappa > 0$ is the magnitude of the phase variation, and $v(t)$ is a zero-mean white Gaussian noise with unity amplitude.

The normalized output photocurrent of the adaptive homodyne detection of the above coherent beam is:
\begin{equation}\label{eq:photocurrent}
\begin{split}
I(t)dt &= 2|\alpha|\sin[\phi(t)-\hat{\phi}(t)]dt + dW(t),\\
&\simeq 2|\alpha|[\phi(t)-\hat{\phi}(t)]dt + dW(t),
\end{split}
\end{equation}
where $\hat{\phi}(t)$ is the phase estimate output of the (feedback) filter, $|\alpha|$ is the coherent amplitude of the beam and $W(t)$ is a Wiener process arising from quantum vacuum fluctuations. Also, here we have taken a linear approximation of the sine function, since the phase estimate would be close to the true phase owing to the feedback. The measurement is appropriately scaled to yield a measurement model as follows:
\begin{equation}
\theta(t) := \frac{I(t)}{2|\alpha|}+\hat{\phi}(t) = \phi(t)+\frac{1}{2|\alpha|}w(t),
\end{equation}
where $w(t) := dW/dt$ is another zero-mean white Gaussian noise with unity amplitude.

Thus we have the following state-space model:
\begin{equation}\label{eq:exact_sys_model}
\begin{split}
\mathsf{Process \, Model:}\quad \dot{\phi}(t) &= -\lambda\phi(t) + \sqrt{\kappa}v(t),\\
\mathsf{Measurement \, Model:}\quad \theta(t) &= \phi(t) + \frac{1}{2|\alpha|}w(t),
\end{split}
\end{equation}
where
\begin{equation}\label{eq:model_uncorrelated}
\begin{split}
\langle v(t)v(\tau)\rangle &= \delta(t-\tau),\\
\langle w(t)w(\tau)\rangle &= \delta(t-\tau),\\
\langle v(t)w(\tau)\rangle &= 0.
\end{split}
\end{equation}

\subsection{Kalman Filter}\label{sec:kalman_filter}
The steady-state Kalman filter for the model (\ref{eq:exact_sys_model}) is obtained from (\ref{eq:filter_form1}), (\ref{eq:filter_form2}), (\ref{eq:kalman_filter_coeffs1}) as:

\vspace*{-3mm}\small
\begin{equation}\label{eq:kalman_filter}
\begin{split}
\dot{\hat{\phi}}(t) &= -\lambda\hat{\phi}(t)+K\left(\theta(t)-\hat{\phi}(t)\right)\\
&= -(\lambda+K)\hat{\phi}(t)+K\phi(t)+\frac{K}{2|\alpha|}w(t).
\end{split}
\end{equation}\normalsize
Here, the Kalman gain $K$ is derived from the error covariance $P = \sigma^2$ of the Kalman filter. The algebraic Riccati equation for $P$ from (\ref{eq:kalman_filter_riccati}) simplifies for our system to:
\begin{equation}\label{eq:kalman_riccati}
-2\lambda P - 4|\alpha|^2P^2 + \kappa = 0.
\end{equation}
The stabilising solution $P^{+}$ of (\ref{eq:kalman_riccati}) is found to be:
\begin{equation}\label{eq:kalman_error}
P^{+} = \frac{\kappa}{\lambda+\sqrt{\lambda^2+4\kappa|\alpha|^2}},
\end{equation}
and the Kalman gain $K$ from (\ref{eq:kalman_filter_coeffs2}) is then:
\begin{equation}\label{eq:kalman_gain}
K = -\lambda + \sqrt{\lambda^2+4\kappa|\alpha|^2}.
\end{equation}
The transfer function (\ref{eq:filter_tf}) of the Kalman filter here is:
\begin{equation}\label{eq:kalman_tf}
G_K(s) := \frac{\hat{\phi}(s)}{\theta(s)} = \frac{K}{s+\lambda+K}.
\end{equation}\vspace*{-3mm}
Equations (\ref{eq:kalman_error}), (\ref{eq:kalman_gain}), (\ref{eq:kalman_filter}) are key equations.

\section{Robust Filter}\label{sec:rob_filter}
In this section, we construct a robust guaranteed-cost filter of the form defined in Ref.~\cite{PM}, as outlined in Appendix \ref{sec:met_grtd_cost}, corresponding to the optimal filter discussed earlier. First, we define our uncertain model.

\subsection{Uncertain Model}
We consider uncertainty in the parameter $\lambda$ as follows:
\begin{equation}
\lambda \to \lambda + \mu\delta\lambda,
\end{equation}
where $|\delta|\leq 1$ is unknown, and $\mu \in [0,1)$ is the level of uncertainty in the model.

Then, the model (\ref{eq:exact_sys_model}) for the uncertain case is:

\small\vspace*{-2mm}
\begin{equation}\label{eq:unc_sys_model}
\begin{split}
\mathsf{Process \, Model:} \, \dot{\phi}(t) &= -(\lambda+\mu\delta\lambda)\phi(t) + \sqrt{\kappa}v(t),\\
\mathsf{Measurement \, Model:} \, \theta(t) &= \phi(t) + \frac{1}{2|\alpha|}w(t).
\end{split}
\end{equation}
\normalsize

\subsection{Guaranteed-Cost Filter}\label{sec:gc_filter}
The steady-state robust filter is then obtained by solving the following Riccati equation for the model (\ref{eq:unc_sys_model}):
\begin{equation}\label{eq:robust_riccati}
-\left(4|\alpha|^2-\epsilon\lambda^2\right)Q^2 - 2\lambda Q + \left(\frac{\mu^2}{\epsilon}+\kappa\right) = 0,
\end{equation}
which is obtained from (\ref{eq:met_riccati2_grtd_cost}) in Theorem \ref{thm:guar_cost_filter} from Appendix \ref{sec:met_grtd_cost}. Here $Q$ is the upper-bound for the error covariance of the robust filter. Also, $\epsilon$ is a positive constant, with additional conditions on its value given in Theorem \ref{thm:guar_cost_filter} in Appendix \ref{sec:met_grtd_cost}.

The stabilising solution $Q^{+}$ of the above equation (\ref{eq:robust_riccati}) is found to be:
\begin{equation}\label{eq:robust_error_bound}
Q^{+} = \frac{\frac{\mu^2}{\epsilon}+\kappa}{\lambda+\sqrt{\lambda^2+\left(4|\alpha|^2- \epsilon\lambda^2\right)\left(\frac{\mu^2}{\epsilon}+\kappa\right)}}.
\end{equation}
The optimum value of $\epsilon$ for which the above bound (\ref{eq:robust_error_bound}) is minimum is obtained as:
\begin{equation}\label{eq:epsilon_opt}
\epsilon_{\rm opt} = \frac{\lambda\mu(1-\mu)+\mu\sqrt{\lambda^2(1-\mu)^2+4\kappa|\alpha|^2}}{\kappa\lambda}.
\end{equation}
Substituting this value of $\epsilon$ in (\ref{eq:robust_error_bound}) yields:
\begin{equation}\label{eq:robust_worst_error}
Q^{+} =\frac{\kappa}{\lambda(1-\mu)+\sqrt{\lambda^2(1-\mu)^2+4\kappa|\alpha|^2}}.
\end{equation}
The robust filter equation is then given as:
\begin{equation}\label{eq:robust_filter}
\begin{split}
\dot{\hat{\phi}}(t) &= \left(-\lambda+\epsilon_{\rm opt} Q^{+}\lambda^2\right)\hat{\phi}(t)\\
&\quad+ 4|\alpha|^2Q^{+}\left(\theta(t)-\hat{\phi}(t)\right)\\
&= \left(-\lambda(1-\mu)-4|\alpha|^2Q^{+}\right)\hat{\phi}(t)\\
&\quad+ 4|\alpha|^2Q^{+}\phi(t)+ 2|\alpha|Q^{+}w(t).
\end{split}
\end{equation}
Note that this does not depend on $\delta$ which is the unknown quantity, so the filter can be used. Equations (\ref{eq:robust_worst_error}), (\ref{eq:robust_filter}) are two of the main results in this paper. The transfer function of the robust filter is
\begin{equation}\label{eq:robust_tf}
G_R(s) := \frac{\hat{\phi}(s)}{\theta(s)} = \frac{4|\alpha|^2Q^{+}}{s+\lambda(1-\mu)+4|\alpha|^2Q^{+}}.
\end{equation}

Clearly, when $\mu = 0$, we have $\epsilon_{\rm opt} = 0$, so that with no uncertainty in the system, (\ref{eq:robust_error_bound}), (\ref{eq:robust_worst_error}) reduce to (\ref{eq:kalman_error}). In other words, (\ref{eq:robust_worst_error}) is the optimal mean-square error for the case where $\lambda$ is known and $\lambda$ is replaced with $\lambda(1-\mu)$. Likewise, with $\mu=0$, the robust filter equation (\ref{eq:robust_filter}) reduces to the Kalman filter equation (\ref{eq:kalman_filter}), and the robust filter transfer function (\ref{eq:robust_tf}) reduces to the Kalman filter transfer function (\ref{eq:kalman_tf}).

Note that the robust filter (\ref{eq:robust_filter}), (\ref{eq:robust_tf}) is the same as the Kalman filter with $\lambda$ replaced by $\lambda(1-\mu)$. Essentially, a robust filter and a Kalman filter are two different filter design algorithms, i.e.~a Kalman filter is designed to be optimal with respect to a certain fixed model and a robust filter is designed to provide acceptable performance for an uncertain system model. However, there is no reason why the Kalman filter and the robust filter design methods should not lead to the same filter in certain situations, even though they are for different models.

\section{Filter Mean-Square Errors for Uncertain System}\label{sec:lyapunov}
In the previous sections we considered the Kalman filter and its error for $\delta=0$ and the robust filter's maximum error. In this section we consider the mean-square errors for the robust and Kalman filters as functions of the unknown parameter $\delta$.  We employ a Lyapunov equation \cite{RPH1} to obtain the desired results.

The uncertain process model from (\ref{eq:unc_sys_model}) is augmented with the filter equation, i.e.~(\ref{eq:kalman_filter}) for the Kalman filter or (\ref{eq:robust_filter}) for the robust filter, to yield the following state-space model:
\begin{equation}
\dot{\overline{\phi}}(t) = \overline{A}\overline{\phi}(t) + \overline{B}\overline{w}(t),
\end{equation}
where $\overline{\phi}(t) := \left[\begin{array}{c}
\phi(t)\\
\hat{\phi}(t)
\end{array}\right]$ and $\overline{w}(t) := \left[\begin{array}{c}
v(t)\\
w(t)
\end{array}\right]$.

Here,
\begin{equation}
\overline{A} = \left[\begin{array}{cc}
-(\lambda+\mu\delta\lambda) & 0\\
\Omega & -\Lambda
\end{array}\right], \, \overline{B} = \left[\begin{array}{cc}
\sqrt{\kappa} & 0\\
0 & \frac{\Omega}{2|\alpha|}
\end{array}\right].
\end{equation}

In the case of the Kalman filter,
\small
\begin{align}
\Omega &= K = -\lambda+\sqrt{\lambda^2+4\kappa|\alpha|^2},\\
\Lambda &= \lambda+K=\sqrt{\lambda^2+4\kappa|\alpha|^2}.
\end{align}
\normalsize

In the case of the robust filter,
\small
\begin{align}
\Omega &= 4|\alpha|^2Q^{+} = -\lambda(1-\mu)+\sqrt{\lambda^2(1-\mu)^2+4\kappa|\alpha|^2},\\
\Lambda &= \lambda(1-\mu)+4|\alpha|^2Q^{+} = \sqrt{\lambda^2(1-\mu)^2+4\kappa|\alpha|^2}.
\end{align}
\normalsize

The steady-state state covariance matrix $P_S$ is obtained by solving a Lyapunov equation as follows:
\begin{equation}\label{eq:lyapunov}
\overline{A}P_S+P_S\overline{A}^T+\overline{B}\overline{B}^T = 0.
\end{equation}

Here, $P_S$ is the symmetric matrix
\begin{equation}
P_S = \langle\overline{\phi}(t)\overline{\phi}^T(t)\rangle = \left[\begin{array}{cc}
P_1 & P_2\\
P_2 & P_3
\end{array}\right],
\end{equation}
where $P_1 = \langle\phi^2(t)\rangle$, $P_2 = \langle\phi(t)\hat{\phi}(t)\rangle$, and $P_3 = \langle\hat{\phi}^2(t)\rangle$.

Then, the estimation error can be written as:
\begin{equation}
\varepsilon(t) = \phi(t) - \hat{\phi}(t) = \left[\begin{array}{cc}
1 & -1
\end{array}\right]\overline{\phi}(t),
\end{equation}
which is of mean zero, since all the quantities determining $\varepsilon(t)$ are of mean zero.

The error covariance is then given as:
\begin{equation}\label{eq:err_cov}
\sigma^2 = \langle\varepsilon^2(t)\rangle = P_1 - 2P_2 + P_3.
\end{equation}

Upon solving (\ref{eq:lyapunov}), we get the following:
\small
\begin{equation}\label{eq:kalman_state_cov}
\begin{split}
P_1 &= \frac{\kappa}{2(\lambda+\mu\delta\lambda)},\\
P_2 &= \frac{\Omega\kappa}{2(\lambda+\mu\delta\lambda)(\Lambda+\lambda+\mu\delta\lambda)},\\
P_3 &= \frac{\Omega^2}{8|\alpha|^2\Lambda}+\frac{\Omega^2\kappa}{2\Lambda(\lambda+\mu\delta\lambda)(\Lambda+\lambda+\mu\delta\lambda)}.
\end{split}
\end{equation}
\normalsize

Substituting (\ref{eq:kalman_state_cov}) in (\ref{eq:err_cov}) and simplifying, we get:
\small
\begin{equation}\label{eq:kalman_error_worst1}
\begin{split}
\sigma^2 &= \kappa\frac{\lambda+\Lambda(1+\mu\delta)}{2\Lambda(1+\mu\delta)(\Lambda+\lambda+\mu\delta\lambda)}+\frac{\Omega^2}{8|\alpha|^2\Lambda}.
\end{split}
\end{equation}
\normalsize

Thus, the mean-square error $\sigma_K^2$ of the Kalman filter for the uncertain system is:
\begin{equation}\label{eq:kalman_error_worst}
\begin{split}
\sigma_K^2 &= \kappa\frac{\lambda+(K+\lambda)(1+\mu\delta)}{2(1+\mu\delta)(K+\lambda)(K+2\lambda+\mu\delta\lambda)}\\
&\quad+\frac{K^2}{8|\alpha|^2(K+\lambda)},
\end{split}
\end{equation}
and the mean-square error $\sigma_R^2$ of the robust filter for the uncertain system is:
\begin{equation}\label{eq:robust_error_worst}
\sigma_R^2 = \kappa\frac{\lambda(1-\mu)^2+J(1+\mu\delta)}{2J(1+\mu\delta)\left(\lambda+\mu\delta\lambda+J\right)}+\frac{\left(4|\alpha|^2Q^{+}\right)^2}{8|\alpha|^2J},
\end{equation}
where $J := \lambda(1-\mu)+4|\alpha|^2Q^{+}$. Equations (\ref{eq:kalman_error_worst}), (\ref{eq:robust_error_worst}) constitute two more main equations in this paper.

\section{Comparison of the Mean-Square Errors of the Filters}\label{sec:ou_comp_errors}
\begin{figure}[!t]
\centering
\includegraphics[width=0.45\textwidth]{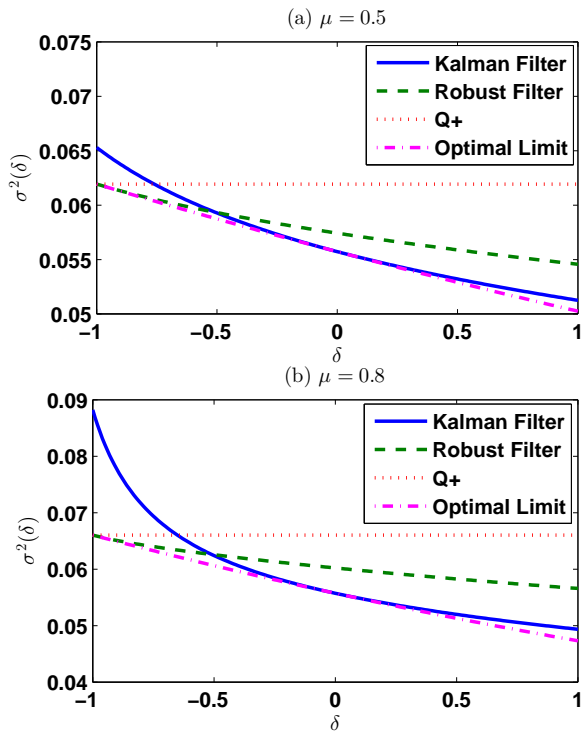}
\caption{\footnotesize Comparison of the mean-square errors of the Kalman and Robust filters for (a) $\mu = 0.5$, (b) $\mu = 0.8$.}
\label{fig:ou_gc_delta}
\end{figure}

The mean-square errors of the Kalman and robust filters can be computed using the technique mentioned in the previous section for the uncertain system for $-1\leq\delta\leq 1$ and plotted on the same graph for comparison. We used the following values for the various parameters for consistency with Ref.~\cite{YNW}: $\lambda = 5.9 \times 10^4$ rad/s, $\kappa = 1.9 \times 10^4$ rad/s and $|\alpha|^2 = 1 \times 10^6$ $\mathrm{s}^{-1}$. Figure \ref{fig:ou_gc_delta} shows the comparisons for $\mu = 0.5$ and $\mu = 0.8$. Also shown on the plots are the upper bounds $Q^{+}$ of the robust filter mean-square error. Moreover, the \emph{optimal limit} for arbitrary $\delta$ (i.e.~the minimum achievable mean-square error if the exact value of $\lambda$ was known at each $\delta$) is found to be (as in Eq. (\ref{eq:kalman_error})):

\small
\begin{equation}\label{eq:unc_opt_error}
\sigma^2_{\rm opt} = \frac{1}{4|\alpha|^2}\left(-(\lambda+\mu\delta\lambda)+\sqrt{(\lambda+\mu\delta\lambda)^2+4\kappa|\alpha|^2}\right),
\end{equation}
\normalsize
which is always less than or equal to $\sigma^2$.

All the curves (except $Q^{+}$) monotonically decrease with increasing $\delta$. This is obvious from the respective expressions (\ref{eq:kalman_error_worst}), (\ref{eq:robust_error_worst}) and (\ref{eq:unc_opt_error}). The Kalman filter mean-square error is lower than the robust filter mean-square error at $\delta = 0$, because the Kalman filter is optimised for the parameters of the exact model. Also, note that $K+\lambda = \sqrt{\lambda^2+4\kappa|\alpha|^2} \geq \sqrt{\lambda^2(1-\mu)^2+4\kappa|\alpha|^2} = J$. Here, $K+\lambda$ is the corner frequency of the Kalman filter, whereas $J$ is that of the robust filter. The corner frequency of the process model from (\ref{eq:unc_sys_model}) is $\lambda(1+\mu\delta)$, which rises with increasing $\delta$. Thus, the Kalman filter outperforms the robust filter with $\delta$ approaching $+1$, owing to the larger corner frequency of the Kalman filter, resulting in less phase information getting truncated compared to the robust filter.

However, as $\delta$ approaches $-1$, the robust filter is superior to the Kalman filter. Since the corner frequency of the process model is now less than $\lambda$, the lower corner frequency of the robust filter provides an advantage. At $\delta = -1$, the worst-case mean-square error $\sigma_w^2 := \sigma^2(\delta=-1,\mu)$ of the robust filter is quantified by $Q^{+}$, that is much lower than that of the Kalman filter. This advantage is greater with $\mu=0.8$ than with $\mu=0.5$. Indeed, as noted earlier, $Q^{+}$ from (\ref{eq:robust_worst_error}) is the optimal mean-square error (\ref{eq:kalman_error}) with $\lambda$ replaced by $\lambda(1-\mu)$ (i.e.~$\lambda+\mu\delta\lambda$, where $\delta$ equals $-1$ and corresponds to the worst-case situation).

\begin{figure}[!b]
\centering
\includegraphics[width=0.45\textwidth]{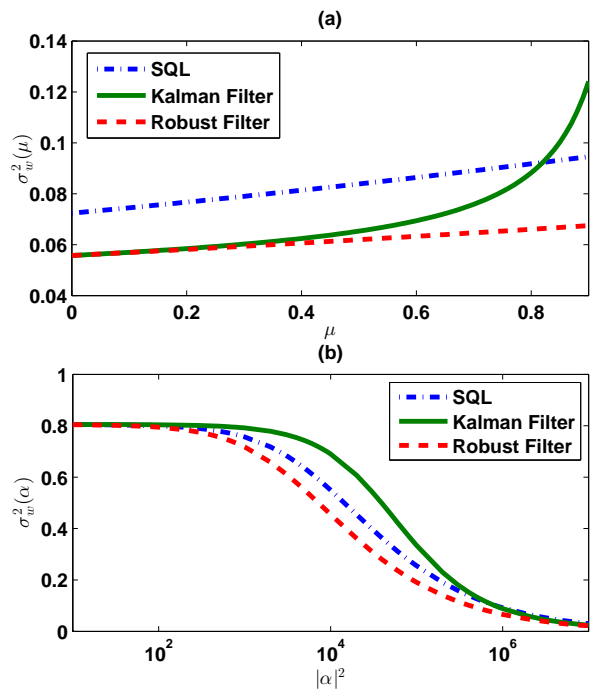}
\caption{\footnotesize Comparison of the worst-case mean-square errors of the Kalman and Robust filters as a function of (a) $\mu$, and (b) $|\alpha|^2$. For (b), we fix $\mu=0.8$.}
\label{fig:ou_gc_worst_mu_alpha}
\end{figure}

The fact that the relative guaranteed worst-case performance of the robust filter with respect to the optimal filter improves with increasing level of uncertainty $\mu$ in the model is illustrated in Fig.~\ref{fig:ou_gc_worst_mu_alpha}-(a). We have also shown the worst-case SQL, which is computed for each $\mu$ as in Appendix \ref{sec:ou_sql}. Note from the plot that the worst-case robust filter mean-square error beats the SQL for all values of $\mu$, whereas the Kalman filter worst-case mean-square error increasingly exceeds the SQL at higher values of $\mu$. Indeed, the worst-case corner frequency of the uncertain process is $\lambda(1-\mu)$, which goes down with increasing value of $\mu$, while the corner frequency of the Kalman filter remains unchanged at $\sqrt{\lambda^2+4\kappa|\alpha|^2}\geq\lambda(1-\mu)$, allowing for less noise getting filtered out with rising $\mu$. Consequently, the mean-square error of the Kalman filter keeps rising with increasing $\mu$ to exceed the SQL. On the other hand, the corner frequency $\sqrt{\lambda^2(1-\mu)^2+4\kappa|\alpha|^2}$ of the robust filter consistently falls with rising $\mu$, allowing for the mean-square error to stay below the SQL throughout.

Figure \ref{fig:ou_gc_worst_mu_alpha}-(b) shows a comparison of the worst-case mean-square errors of the two filters as a function of the photon flux $|\alpha|^2$ with the uncertainty level set at $\mu = 0.8$. We also plot the SQL for comparison. We see that our robust filter exhibits an optimal photon number for which its worst-case performance is the best relative to the Kalman filter. This is similar to the observation made in Ref.~\cite{RPH1} (that the robust smoother admitted an optimal photon number when compared to the optimal smoother). Also, the worst-case mean-square error of the robust filter is always lower than the SQL, while that of the Kalman filter is not. Indeed, for low values of photon flux, the corner frequency $\sqrt{\lambda^2+4\kappa|\alpha|^2}$ of the Kalman filter is well beyond the corner frequency $\lambda(1-\mu)$ of the uncertain phase, resulting in its mean-square error being quite high. However, with rising $|\alpha|^2$ beyond the point where $\lambda^2$ is comparable to $4\kappa|\alpha|^2$, the value of $\lambda$ (that remains fixed here) becomes increasingly less dominant and therefore, the impact of any uncertainty in $\lambda$ on the estimation precision keeps reducing. The behavior of the robust filter is similar, although its corner frequency $\sqrt{\lambda^2(1-\mu)^2+4\kappa|\alpha|^2}$ clearly stays much lower than that of the Kalman filter, resulting in lower mean-square errors. Also, the mean-square error of the robust filter starts falling earlier compared to the Kalman filter with rising $|\alpha|^2$ at the point where $\lambda^2(1-\mu)^2$ is comparable to $4\kappa|\alpha|^2$. A similar observation can be made for the SQL too, the worst-case corner frequency for which is $\lambda(1-\mu)+K_{SQL}=\sqrt{\lambda^2(1-\mu)^2+2\kappa|\alpha|^2}$ from (\ref{eq:ou_sql_tf}).

All these plots indicate that the worst-case mean-square error, quantified by $Q^{+}$, of our robust filter is guaranteed to beat that of the Kalman filter for all situations. Moreover, the robust filter worst-case mean-square error beats the SQL, even when the Kalman filter worst-case mean-square error does not. So if the SQL is considered as the tolerable threshold of the mean-square error for our system, our robust filter guarantees that the mean-square error never exceeds this threshold. This is exactly why robust design methods are powerful for practical engineering applications (as also noted in Ref.~\cite{RPH1}).

\section{Error Analysis}\label{sec:error_analysis}
In this section, we explicitly prove that the robust filter mean-square error is guaranteed to beat the Kalman filter mean-square error and the SQL in the worst-case situation for all admissible values of $\lambda$, $\kappa$, $|\alpha|$ and $\mu$.

\paragraph{Robust filter vs.~Kalman filter:}
Here, we intend to show that the worst-case Kalman filter mean-square error, i.e.~(\ref{eq:kalman_error_worst}) with $\delta = -1$, is always beaten by the worst-case robust filter mean-square error, given by $Q^{+}$ from (\ref{eq:robust_worst_error}). As mentioned earlier, it is immediately obvious from (\ref{eq:robust_worst_error}), when compared to (\ref{eq:kalman_error}), that $Q^{+}$ is the optimal mean-square error at $\delta=-1$. Therefore, it can be no larger than the mean-square error at this value of $\delta$ of the Kalman filter, which is constructed to be optimal for the exact model having $\delta=0$ and is suboptimal everywhere else including at $\delta=-1$.

\paragraph{Robust filter vs.~SQL:}
Here, we intend to show that the worst-case robust filter error covariance, quantified by $Q^{+}$ from (\ref{eq:robust_worst_error}), always beats the SQL, given by (\ref{eq:ou_sql}) with $\delta = -1$, for all $\lambda,\kappa,|\alpha|\geq 0$ and $0\leq\mu<1$.

Subtracting (\ref{eq:robust_worst_error}) from (\ref{eq:ou_sql}) and simplifying with $\delta=-1$, we get:
\small
\begin{equation}
\begin{split}
P_{SQL}(\delta&=-1)-Q^{+}\\
&=\frac{-\lambda(1-\mu)+\sqrt{\lambda^2(1-\mu)^2+2\kappa|\alpha|^2}}{2|\alpha|^2}\\
&\quad-\frac{-\lambda(1-\mu)+\sqrt{\lambda^2(1-\mu)^2+4\kappa|\alpha|^2}}{4|\alpha|^2}.
\end{split}
\end{equation}
\normalsize

One can see that in order for the above quantity to be greater than or equal to zero, we need to have:
\small
\begin{equation*}
2\sqrt{\lambda^2(1-\mu)^2+2\kappa|\alpha|^2}-\lambda(1-\mu)\geq\sqrt{\lambda^2(1-\mu)^2+4\kappa|\alpha|^2},
\end{equation*}
\normalsize
which is proven in Appendix \ref{sec:proof_rob_sql}.

Hence, $P_{SQL}(\delta=-1)-Q^{+}\geq 0$, i.e.~the robust filter mean-square error for the uncertain system is always less than or equal to the SQL in the worst-case situation. This is a key result in our paper.

\section{Effective Quantum Efficiency}\label{sec:ou_eff_qeff}
In all of the above, we have assumed perfect detector efficiency. That is, the photon flux $|\alpha|^2$ equals the square of the absolute amplitude of the input coherent state $\alpha_c e^{i\phi(t)}$. In practice, the photon flux is:
\begin{equation}
|\alpha|^2 = \eta_d|\alpha_c|^2,
\end{equation}
where $0 < \eta_d \leq 1$ is the homodyne detector efficiency.

\begin{figure}[!b]
\centering
\includegraphics[width=0.45\textwidth]{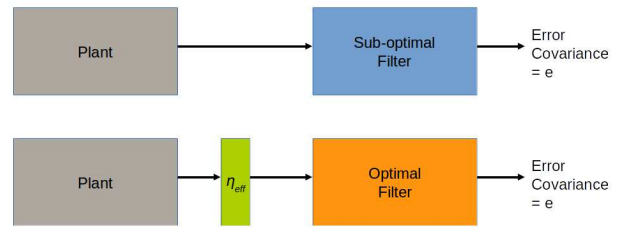}
\caption{\footnotesize A suboptimal estimation process may yield the same mean-square error as the optimal estimation process preceded by a loss.}
\label{fig:filter1}
\end{figure}

The inefficiency of detectors increases the mean-square error, just as using a suboptimal filter does. This suggests considering an \emph{effective quantum efficiency} for our filters. That is, the effective quantum efficiency is that which would increase the mean-square error as much as using the suboptimal filter (see Fig.~\ref{fig:filter1}). The effective quantum efficiency $\eta_{\rm eff}$ for a suboptimal filter is defined to be that such that the optimal filter with this efficiency gives the same mean-square error as the suboptimal filter with unit efficiency. This effective efficiency is inspired by that in Ref.~\cite{HBL}, where a measurement scheme is mathematically identical to a combination of loss followed by a better measurement. The difference is that here we are just matching the mean-square error.

Consider the plant here to be our uncertain model from (\ref{eq:unc_sys_model}). Also, suppose a suboptimal filter yields mean-square error $e$ as shown in the figure. Then, the mean-square error yielded by the optimal (Kalman) filter preceded by a loss introduced in the form of $\eta_{\rm eff}$ as in the figure will be equal to $e$ if we have
\begin{equation}\label{eq:eff_eqn_solve}
e = \frac{1}{4\eta_{\rm eff}|\alpha|^2}\left(-\lambda_u+\sqrt{\lambda_u^2+4\kappa\eta_{\rm eff}|\alpha|^2}\right),
\end{equation}
where the right-hand side has been obtained with $\lambda_u := (\lambda+\mu\delta\lambda)$ for the uncertain system (\ref{eq:unc_sys_model}) as (\ref{eq:kalman_error}) was obtained for the exact model (\ref{eq:exact_sys_model}). Upon solving (\ref{eq:eff_eqn_solve}) for $\eta_{\rm eff}$, we get:
\begin{equation}\label{eq:eff_quantum_efficiency}
\eta_{\rm eff} = \frac{\kappa-2e(\lambda+\mu\delta\lambda)}{4|\alpha|^2e^2}.
\end{equation}
This equation is one of the main results in this paper.

Now, consider Fig.~\ref{fig:ou_gc_delta} and note that the filters are suboptimal everywhere except the Kalman filter at $\delta = 0$ and the robust filter at $\delta = -1$. Then, the mean-square errors $\sigma_K^2$ from (\ref{eq:kalman_error_worst}) or $\sigma_R^2$ from (\ref{eq:robust_error_worst}) for these filters (that were plotted in Fig.~\ref{fig:ou_gc_delta}) may be obtained by an effective filter, involving an optimal Kalman filter preceded by a loss channel with transmissivity $\eta_{\rm eff}$ given by (\ref{eq:eff_quantum_efficiency}) with $e=\sigma_K^2$ or $\sigma_R^2$, respectively.

At this point, let us ensure that $\eta_{\rm eff}$ from (\ref{eq:eff_quantum_efficiency}) satisfies the lower and upper bounds of $0$ and $1$, respectively, as desired. Note from (\ref{eq:eff_quantum_efficiency}), that we will have $\eta_{\rm eff} > 0$, if
\begin{equation}\label{eq:eff_qeff_lower}
e < \frac{\kappa}{2(\lambda+\mu\delta\lambda)}.
\end{equation}

The right-hand side above is $P_1$ from (\ref{eq:kalman_state_cov}). This $P_1$ is simply the state covariance $\langle\phi^2(t)\rangle$ for the uncertain process model in (\ref{eq:unc_sys_model}). On the other hand, $e$ is the error covariance $\langle(\phi(t)-\hat{\phi}(t))^2\rangle$. Note that the right-hand side of (\ref{eq:eff_qeff_lower}) equals the error covariance if the phase estimate is zero. In most practical situations, the phase estimate would be better than this, giving a positive efficiency.

Next, note from (\ref{eq:eff_quantum_efficiency}), that we will have $\eta_{\rm eff} \leq 1$, if
\begin{equation}\label{eq:eff_qeff_upper}
\begin{split}
&\quad4|\alpha|^2e^2 + 2(\lambda+\mu\delta\lambda)e - \kappa \geq 0\\
\Rightarrow &\left(e+\frac{(\lambda+\mu\delta\lambda)- \sqrt{(\lambda+\mu\delta\lambda)^2+4\kappa|\alpha|^2}}{4|\alpha|^2}\right)\\
&\times\left(e+\frac{(\lambda+\mu\delta\lambda)+\sqrt{(\lambda+\mu\delta\lambda)^2+4\kappa|\alpha|^2}}{4|\alpha|^2}\right)\\
&\geq 0.
\end{split}
\end{equation}

Clearly, the second term in the product on the left-hand side above is positive. The first term in the product is merely $e-\sigma_{\rm opt}^2$, which is evidently greater than or equal to zero, since the estimation mean-square error of any filter is greater than or equal to that of the optimal filter. Hence, (\ref{eq:eff_qeff_upper}) is satisfied.

\begin{figure}[!t]
\centering
\includegraphics[width=0.42\textwidth]{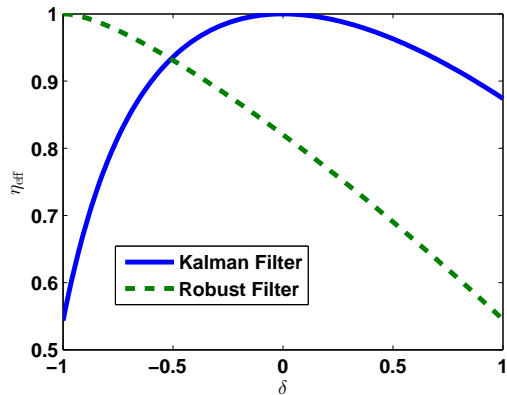}
\caption{\footnotesize Effective Quantum Efficiency: Comparison of the effective quantum efficiencies of the Kalman and Robust filters for the uncertain system as a function of $\delta$ with $\mu=0.8$.}
\label{fig:ou_gc_delta_eff_qeff_delta}
\end{figure}

Now, it is interesting to see how the effective quantum efficiency $\eta_{\rm eff}$ from (\ref{eq:eff_quantum_efficiency}) varies with various parameters for our Kalman and robust filters for the uncertain system. Figure \ref{fig:ou_gc_delta_eff_qeff_delta} shows a comparison of the effective quantum efficiencies of the filters as a function of $\delta$ for $\mu=0.8$. The efficiency $\eta_{\rm eff}$ for the robust filter is unity in the worst-error case, i.e.~at $\delta=-1$, since the robust filter is optimal for this value of $\delta$. On the other hand, $\eta_{\rm eff}$ for the Kalman filter is far below unity in the worst-error case. Also, note that $\eta_{\rm eff}=1$ for the Kalman filter at $\delta=0$, since the Kalman filter is optimal for the exact model, which corresponds to $\delta=0$.

\begin{figure}[!t]
\centering
\includegraphics[width=0.45\textwidth]{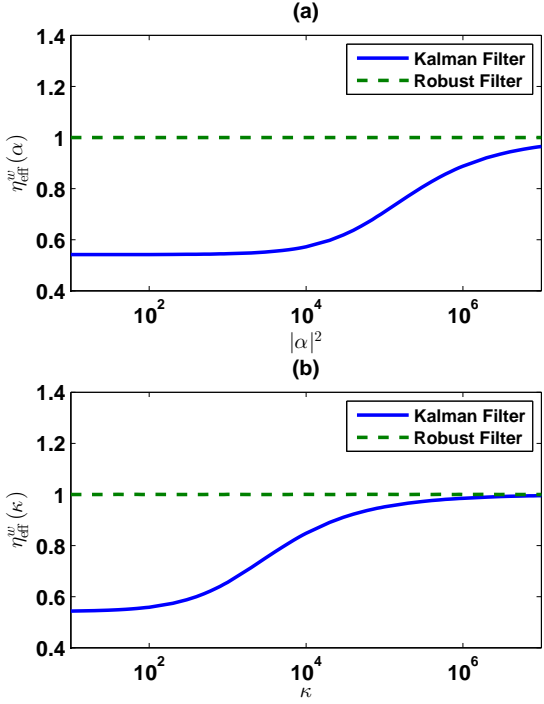}
\caption{\footnotesize Effective Quantum Efficiency: Comparison of the worst-error case effective quantum efficiencies of the Kalman and Robust filters as a function of (a) $|\alpha|^2$, and (b) $\kappa$. Here, we fix $\mu=0.5$.}
\label{fig:ou_gc_worst_eff_qeff_all}
\end{figure}

It is further interesting to note our robust filter's advantage over the Kalman filter in terms of the effective quantum efficiency in the worst-error case situation (i.e.~at $\delta = -1$), that we denote as $\eta_{\rm eff}^w$. Figures \ref{fig:ou_gc_worst_eff_qeff_all}-(a) and (b) show comparisons of $\eta_{\rm eff}^w$ of the filters as functions of $|\alpha|^2$ and $\kappa$, respectively, for $\mu = 0.5$. Since the robust filter is always optimal at $\delta=-1$, its effective quantum efficiency always remains unity in the worst-error case, whereas that of the Kalman filter varies significantly and approaches $\eta_{\rm eff}^w=1$ of the robust filter for increasingly higher values of $\kappa$ and $|\alpha|^2$. Indeed, for low values of $\kappa$ or $|\alpha|^2$, the corner frequency $\sqrt{\lambda^2+4\kappa|\alpha|^2}$ of the Kalman filter is well beyond the worst-case corner frequency $\lambda(1-\mu)$ of the uncertain phase, yielding poor estimates and therefore poor effective quantum efficiency. However, beyond the point at which $4\kappa|\alpha|^2$ becomes comparable to $\lambda^2$, the value of $\lambda$ (that remains fixed here) becomes increasingly less dominant and therefore the impact of any uncertainty in $\lambda$ on the quality of the estimate starts diminishing, resulting in improved effective quantum efficiency. Thus, our robust filter always achieves the maximum effective quantum efficiency that is possible in the worst-case situation, unlike the Kalman filter. This is because, while the robust filter is designed to be optimal in the worst-error case, the Kalman filter is optimal for the exact model without uncertainty.

\section{Effective Noise Power}\label{sec:ou_eff_np}
In Section \ref{sec:ou_eff_qeff}, we defined effective quantum efficiency for a suboptimal measurement with respect to the optimal measurement. However, the effect of the loss is an increase in the error covariance, that can be viewed as resulting from a scaling of the phase (the same model involving the same differential equations with suitably scaled variables) and not the suboptimal filter. We illustrate this in the following.

The measurement model equation is
\begin{equation}
\theta(t) = \phi(t) + \frac{1}{2|\alpha|}w(t).
\end{equation}
With efficiency $\eta$ we have
\begin{equation}
\theta(t) = \phi(t) + \frac{1}{2\sqrt{\eta}|\alpha|}w(t).
\end{equation}

Now let us scale the time to $q=\sqrt{\eta} t$. Then, in terms of the new time we have a new Gaussian noise $w_1(q) = \eta^{1/4}w(t)$, which gives
\begin{equation}
\theta(q) = \phi(q) + \frac{1}{2\eta^{1/4}|\alpha|}w_1(q),
\end{equation}
or
\begin{equation}
\eta^{1/4}\theta(q) = \eta^{1/4}\phi(q) + \frac{1}{2|\alpha|}w_1(q).
\end{equation}
Let us define $\theta_1(q):=\eta^{1/4}\theta(q)$ and $\phi_1(q):=\eta^{1/4}\phi(q)$. In terms of these new quantities the equation is
\begin{equation}
\theta_1(q) = \phi_1(q) + \frac{1}{2|\alpha|}w_1(q).
\end{equation}
The modified equation for $\phi$ is
\begin{equation}
\eta^{1/2}\dot{\phi}(q) = -\lambda \phi(q) + \eta^{1/4}\sqrt{\kappa}v_1(q),
\end{equation}
or
\begin{equation}
\dot{\phi}_1(q) = - \eta^{-1/2} \lambda \phi_1(q) + \sqrt{\kappa}v_1(q).
\end{equation}
Defining $\lambda_1 := \eta^{-1/2} \lambda$, we get
\begin{equation}
\dot{\phi}_1(q) = - \lambda_1\phi_1(q) + \sqrt{\kappa}v_1(q).
\end{equation}
That is, the loss on $|\alpha|$ gives the same plant but with a scaled time, a scaled $\lambda$ and scaled phases.

Now consider the filter
\begin{equation}
\dot{\hat{\phi}}(t) = -\lambda\hat{\phi}(t) + K(\theta(t)-\hat{\phi}(t)).
\end{equation}
In terms of the scaled time
\begin{equation}
\eta^{1/2}\dot{\hat{\phi}}(q) = -\lambda\hat{\phi}(q) + K(\theta(q)-\hat{\phi}(q)).
\end{equation}
Considering scaled variable as well and transferring the $\eta$ to the other side,
\small
\begin{equation}
\dot{\hat{\phi}}_1(q) = -(\eta^{-1/2}\lambda)\hat{\phi}_1(q) + (\eta^{-1/2}K)(\theta_1(q)-\hat{\phi}_1(q)).
\end{equation}
\normalsize
Let us define $K_1:=\eta^{-1/2}K$, so this equation becomes
\begin{equation}
\dot{\hat{\phi}}_1(q) = -\lambda_1\hat{\phi}_1(q) + K_1(\theta_1(q)-\hat{\phi}_1(q)).
\end{equation}

If the Kalman gain was previously calculated for coherent amplitude of $\eta^{1/2}\alpha$, then $K=-\lambda+\sqrt{\lambda^2+4\kappa\eta|\alpha|^2}$. That means that $K_1:=\eta^{-1/2}K = -\lambda_1+\sqrt{\lambda_1^2+4\kappa|\alpha|^2}$. This, in turn, means that $K_1$ is the desired Kalman gain for the state with scaled $\lambda$ without loss. The error covariance is therefore just multiplied by the scaling factor between $\phi_1$ and $\phi$, and is therefore $\eta^{-1/2}$ times the error covariance without loss. Thus, the effect of the loss can be absorbed into scaling of the various variables to give the same differential equations as before, but still with the optimal Kalman filter (and not the suboptimal filter). In other words, the loss is equivalent to having a phase that varies by a larger amount, which would be like adding some noise and trying to measure the joint phase variation plus noise accurately.
\begin{figure}[!b]
\includegraphics[width=0.45\textwidth]{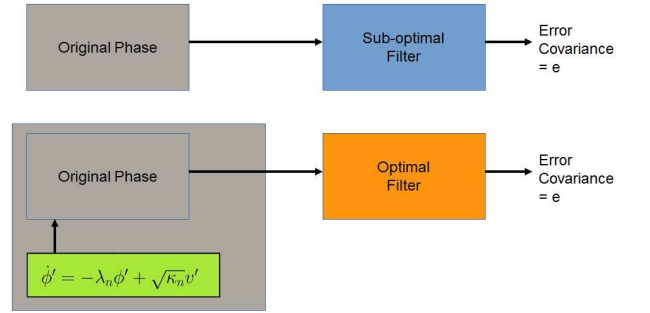}
\caption{\footnotesize A suboptimal estimation process of a phase may yield the same mean-square error as the optimal estimation process of the phase plus additional noise.}
\label{fig:filter2}
\end{figure}

Here, we intend to define an \emph{effective noise power} for the OU phase under measurement, motivated by the result that a suboptimal measurement of a constant phase is equivalent to an optimal measurement on a different state \cite{MJWH1,MJWH2}. In some cases the different state can correspond to the original state with added phase noise. Here, we instead consider added phase noise with an optimal measurement for the varying phase case. We consider an additional OU noise $\dot{\phi}' := -\lambda_n\phi' +\sqrt{\kappa_n}v'$, added to the original phase $\phi$. This is depicted schematically in Fig.~\ref{fig:filter2}. We measure $\phi+\phi'$, but estimate only $\phi$ and not $\phi+\phi'$. This is distinct from trying to estimate $\phi+\phi'$, which is directly equivalent to a scaled scheme of measurement with the optimal filter (similar to the case with effective quantum efficiency discussed above).

The model can then be given in state-space form as:
\begin{equation}\label{eq:add_ou_model}
\begin{split}
\dot{\overline{\phi}} &= A\overline{\phi} + B\overline{v},\\
\theta' &= C\overline{\phi} + Dw,\\
z &= L\overline{\phi},
\end{split}
\end{equation}
where $z$ determines the parameter to be estimated, and
\begin{equation}\label{eq:add_ou_matrices}
\begin{split}
\overline{\phi} &:= \left[\begin{array}{c}
\phi\\
\phi'
\end{array}\right], \quad \overline{v} := \left[\begin{array}{c}
v\\
v'
\end{array}\right],\\
A &= \left[\begin{array}{cc}
-\lambda_u & 0\\
0 & -\lambda_n
\end{array}\right], \quad B = \left[\begin{array}{cc}
\sqrt{\kappa} & 0\\
0 & \sqrt{\kappa_n}
\end{array}\right],\\
C &= \left[\begin{array}{cc}
1 & 1
\end{array}\right], \quad D = \frac{1}{2|\alpha|}, \quad L = \left[\begin{array}{cc}
1 & 0
\end{array}\right].
\end{split}
\end{equation}
Here, $\lambda_u = \lambda+\mu\delta\lambda$ and $v$, $v'$ and $w$ are mutually uncorrelated Gaussian white noises, each with unity amplitude. Also, we take $\lambda_n = \lambda_u$ for simplicity. Note that while the new measurement $\theta'$ considers the measurement of the combination $\phi+\phi'$ (through $C$), the variable to be estimated $z$ is just the original phase $\phi$ (through $L$). Then, if the optimal steady-state estimate is given by $\hat{z}$, the optimal mean-square estimation error $\langle (z-\hat{z})^2\rangle$ is obtained by solving the following Kalman filter algebraic Riccati equation:
\begin{equation}\label{eq:add_ou_riccati}
AP + PA^T - PC^T(DD^T)^{-1}CP + BB^T = 0,
\end{equation}
where $P$ is the symmetric matrix:
\begin{equation}
P := \left[\begin{array}{cc}
P_1 & P_2\\
P_2 & P_3
\end{array}\right],
\end{equation}
such that the desired mean-square estimation error is:
\begin{equation}\label{eq:add_ou_est_error}
\langle(z-\hat{z})^2\rangle = LPL^T = P_1.
\end{equation}

Now, expanding (\ref{eq:add_ou_riccati}), we get the following equations:
\begin{align}
-2\lambda_uP_1 + \kappa - 4|\alpha|^2(P_1+P_2)^2 &= 0,\label{eq:add_ou_riccati_expand1}\\
-2\lambda_uP_2 - 4|\alpha|^2(P_1+P_2)(P_2+P_3) &= 0,\label{eq:add_ou_riccati_expand2}\\
-2\lambda_uP_3 + \kappa_n - 4|\alpha|^2(P_2+P_3)^2 &= 0.\label{eq:add_ou_riccati_expand3}
\end{align}

Upon solving the system of equations (\ref{eq:add_ou_riccati_expand1})-(\ref{eq:add_ou_riccati_expand3}) for $P_1$, $P_2$ and $P_3$, we get:
\begin{equation}\label{eq:add_ou_err_cov}
\begin{split}
P_1 &= \frac{\kappa\left(2|\alpha|^2\kappa_n(\kappa+\kappa_n) -\kappa\lambda_u^2+\kappa\lambda_u\beta\right)}{4|\alpha|^2(\kappa+\kappa_n)^2\lambda_u},\\
P_2 &= \frac{-2|\alpha|^2\kappa\kappa_n(\kappa+\kappa_n)-\kappa\kappa_n\lambda_u^2 +\kappa\kappa_n\lambda_u\beta}{4|\alpha|^2(\kappa+\kappa_n)^2\lambda_u},\\
P_3 &= \frac{\kappa_n\left(2|\alpha|^2\kappa(\kappa+\kappa_n) -\kappa_n\lambda_u^2+\kappa_n\lambda_u\beta\right)}{4|\alpha|^2(\kappa+\kappa_n)^2\lambda_u},\\
\beta :&= \sqrt{4|\alpha|^2(\kappa+\kappa_n)+\lambda_u^2}.
\end{split}
\end{equation}

Next, we deduce $\kappa_n \geq 0$, such that the optimal filter for (\ref{eq:add_ou_model}) yields a variance (\ref{eq:add_ou_est_error}) equal to the estimation mean-square error $e$ of a given arbitrary filter for the original model (\ref{eq:unc_sys_model}). Thus, we should have $P_1 = e$ and $\kappa_n$ may be derived to be:
\begin{equation}\label{eq:kappa_n}
\kappa_n = \frac{\kappa\lambda_u\left(2|\alpha|e-\sqrt{\kappa-2\lambda_ue}\right)}{|\alpha|(\kappa-2\lambda_ue)}.
\end{equation}

The \emph{effective noise power} of the phase $\phi+\phi'$ being measured is then given by 
\begin{equation}\label{eq:kappa_eff}
\kappa_{\rm eff} := \left[\begin{array}{cc}
\sqrt{\kappa} & \sqrt{\kappa_n}
\end{array}\right]\left[\begin{array}{c}
\sqrt{\kappa}\\
\sqrt{\kappa_n}
\end{array}\right] = \kappa + \kappa_n.
\end{equation}
This is because one may verify that the phase being measured (not estimated) is effectively another OU process as below obtained from (\ref{eq:add_ou_model}):
\begin{equation}
\dot{\phi}+\dot{\phi}' = -\lambda_u(\phi+\phi') + \left[\begin{array}{cc}
\sqrt{\kappa} & \sqrt{\kappa_n}
\end{array}\right]\left[\begin{array}{c}
v\\
v'
\end{array}\right].
\end{equation}
Equations (\ref{eq:kappa_eff}), (\ref{eq:kappa_n}) are two more main results.

Note that in measurement theory for a constant phase, modifying the state suitably and using an optimal measurement gives the same result as a suboptimal measurement \cite{MJWH1,MJWH2}. As a result, it yields the same overall probability distribution for the estimation errors, not just the mean-square error. This means that the measurement can be regarded as being the same in Fig.~\ref{fig:filter2} in a deeper sense than considered so far here by demonstrating that it will reproduce the same probability distribution for the measurement errors. This can be tested for a varying phase, as considered here, by demonstrating the equivalence not just in terms of mean-square errors but also in terms of the two-time error correlations.
\begin{figure}[!t]
\centering
\includegraphics[width=0.45\textwidth]{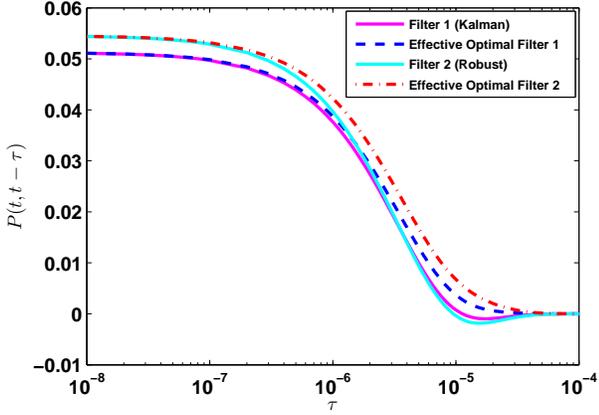}
\caption{\footnotesize Comparison between the two-time error-correlations of various filters as a function of the time-difference $\tau$ for $\mu=0.5$ and $\delta=1$. The various traces in the plot are: (i) The pink solid line is the two-time error correlation at each $\tau$ of the Kalman filter constructed for the exact model and applied to the uncertain system; (ii) The cyan solid line is the two-time error correlation at each $\tau$ of the robust filter constructed for the uncertain model and applied to the uncertain system; (iii) The blue dotted line is the two-time error correlation at each $\tau$ of the optimal filter constructed for the uncertain model plus an added noise corresponding to the Kalman filter of trace (i); and (iv) The red dotted line is the two-time error correlation at each $\tau$ of the optimal filter constructed for the uncertain model plus an added noise corresponding to the robust filter of trace (ii).}
\label{fig:ou_gc_effkappa_2t}
\end{figure}

Figure \ref{fig:ou_gc_effkappa_2t} shows the comparison of the two-time error correlations of the various filters for $\delta = 1$, $\mu=0.5$, $\lambda_n = \lambda_u$ and $\kappa_n$ as derived in (\ref{eq:kappa_n}) and, as used before, $\lambda = 5.9 \times 10^4$ rad/s, $\kappa = 1.9 \times 10^4$ rad/s and $|\alpha|^2 = 1 \times 10^6$ $\mathrm{s}^{-1}$. The two-time error correlations are computed as in Appendix \ref{sec:error_2t}. Clearly, the effective optimal filter we obtained for the Kalman filter is not equivalent to the Kalman filter at this value of $\delta$ in terms of the two-time error correlations; likewise for the robust filter. Thus, we observe that with an added OU noise as considered here, the effective optimal filter is equivalent to the suboptimal filter (our Kalman filter at $\delta \neq 0$ or our robust filter at $\delta \neq -1$) in terms of mean-square estimation errors (this is expected since $\kappa_n$ of the added OU noise is derived such that the mean-square errors match up) only, and not two-time error correlations as well. There are analytic reasons, as presented in Appendix \ref{sec:more_analysis}, that suggest that the two-time error correlations cannot possibly be made to match up. However, for the purposes of our work here, where our focus is to define and compare the effective noise powers for our different estimators, it suffices to use only mean-square estimation errors.
\begin{figure}[!b]
\centering
\includegraphics[width=0.42\textwidth]{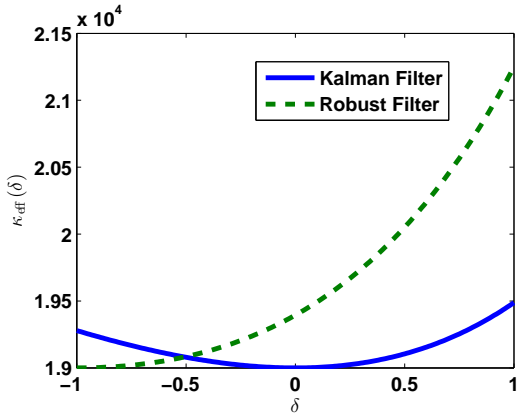}
\caption{\footnotesize Effective Noise Power: Comparison of the effective noise powers of the Kalman and Robust filters for the uncertain system as a function of $\delta$ with $\mu=0.5$.}
\label{fig:ou_gc_delta_effkappa_delta}
\end{figure}

We now show in Fig.~\ref{fig:ou_gc_delta_effkappa_delta} a comparison of the effective noise powers of the filters as a function of $\delta$ for $\mu=0.5$. The effective noise power $\kappa_{\rm eff}$ for the robust filter is at a minimum and equal to $\kappa$ in the worst-error case, i.e.~at $\delta=-1$, since the robust filter is optimal for this value of $\delta$. By contrast, $\kappa_{\rm eff}$ for the Kalman filter is more than that of the robust filter in the worst-error case. Also, note that $\kappa_{\rm eff}$ is at a minimum (i.e.~equal to $\kappa$) for the Kalman filter at $\delta=0$, since the Kalman filter is optimal for the exact model, which corresponds to $\delta=0$.

\begin{figure}[!b]
\centering
\includegraphics[width=0.45\textwidth]{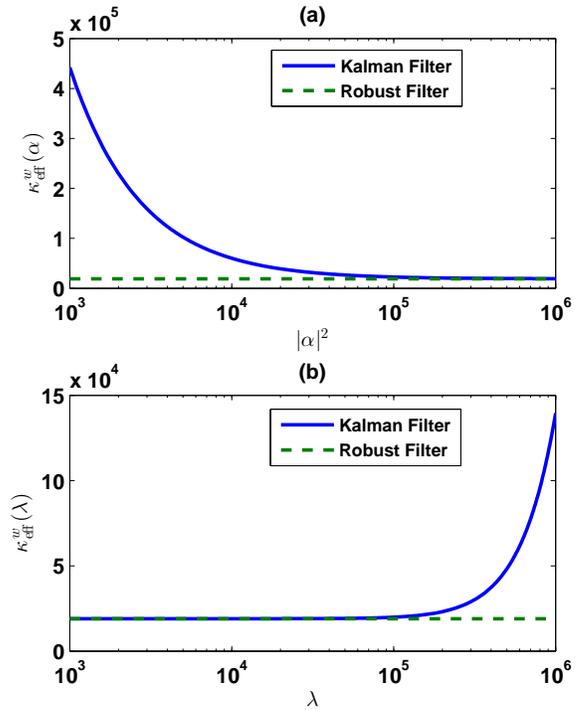}
\caption{\footnotesize Effective Noise Power: Comparison of the worst-error case effective noise powers of the Kalman and Robust filters as a function of (a) $|\alpha|^2$, and (b) $\lambda$. Here, we fix $\mu = 0.5$.}
\label{fig:ou_gc_worst_effkappa_all}
\end{figure}

It is interesting to note our robust filter's advantage over the Kalman filter in terms of the effective noise power in the worst-error case situation (i.e.~at $\delta = -1$), that we denote as $\kappa^w_{\rm eff}$. Figures \ref{fig:ou_gc_worst_effkappa_all}-(a) and (b) show comparisons of $\kappa^w_{\rm eff}$ of the filters as functions of $|\alpha|^2$ and $\lambda$, respectively, with $\mu = 0.5$. We note that the worst-error case effective noise power of the robust filter always remains constant at $\kappa$, while that of the Kalman filter approaches $\kappa^w_{\rm eff} = \kappa$ of the robust filter for higher values of $|\alpha|^2$ and lower values of $\lambda$. Intuitively, since the worst-case mean-square error rises for the Kalman filter with decreasing $|\alpha|^2$, the corresponding effective optimal filter incurs higher noise power. Moreover, with rising $\lambda$, the size of the uncertainty in the model increases, resulting in higher effective noise power for the Kalman filter.

All these plots show that when considering the worst-case scenario, our robust filter always incurs the least effective noise possible, whereas the Kalman filter often effectively suffers from much higher noise due to the uncertainty in the model. This is again because of the fact that the robust filter design involves optimising the error in the worst-case situation, while that for the Kalman filter considers zero uncertainty in the model.

\section{Conclusion}\label{sec:concl}
In this paper, we studied guaranteed-cost robust adaptive quantum phase estimation for a coherent state with parametric uncertainty explicitly introduced in the model in a systematic state-space setting. Our robust filter comes with a guaranteed cost that quantifies the worst-case mean-square estimation error, which is guaranteed to beat that of the Kalman filter and the SQL for the uncertain system. We showed that if the SQL is taken as the tolerable threshold, our robust filter mean-square error always remains under this threshold unlike the Kalman filter. We illustrated that this guaranteed cost allows for our robust filter to always achieve the maximum effective quantum efficiency and the minimum possible effective noise power in the worst-case scenario. In contrast to the robust fixed-interval smoother from Ref.~\cite{RPH1}, our robust filter considered here helps overcome practical challenges facing quantum parameter estimation in applications such as quantum computing, where real-time estimates are of greater interest than more precise estimates.

Moreover, in deriving an effective noise power for our robust and Kalman filters, we matched only the mean-square errors of these filters for the uncertain system with the optimal filter with an added noise. We saw that it does not seem possible to also match up the two-time error correlations. Thus, another key result here is that it appears that the results from Refs.~\cite{MJWH1,MJWH2} for a constant phase do not generalize to the case of a varying phase, such that the overall probability distributions of the estimation errors match up.

\begin{acknowledgments}
This work was supported by the Australian Research Council (ARC) under grants FL110100020 (IRP) and CE110001027 (EHH). SR was funded by the Singapore National Research Foundation Grant No.~NRF-NRFF2011-07 and the Singapore Ministry of Education Academic Research Fund Tier 1 Project R-263-000-C06-112, and is currently funded by the UK National Quantum Technologies Programme (EP/M01326X/1, EP/M013243/1). DWB is funded by an ARC Future Fellowship (FT100100761) and an ARC Discovery Project (DP160102426). This work was also supported by the US Air Force Office of Scientific Research (AFOSR). This material is based on research sponsored by the Air Force Research Laboratory, under agreement numbers FA2386-12-1-4075 and FA2386-16-1-4065 (IRP). The US Government is authorized to reproduce and distribute reprints for Governmental purposes notwithstanding any copyright notation thereon. The views and conclusions contained herein are those of the authors and should not be interpreted as necessarily representing the official policies or endorsements, either expressed or implied, of the Air Force Research Laboratory or the US Government.

Moreover, SR thanks Howard Wiseman, Hongbin Song, Aleksandar Davidovic, Obaid Ur Rehman, Trevor Wheatley, Hidehiro Yonezawa, Mohamed Mabrok and Katanya Kuntz for discussions in relation to this work.
\end{acknowledgments}

\bibliography{rob_gc_biblio}

\vspace*{-5mm}
\appendix
\section{Robust Guaranteed-Cost Filter}\label{sec:met_grtd_cost}
We consider here the theory underlying a steady-state robust estimator for a class of uncertain linear systems with norm-bounded uncertainty \cite{PM}. This extends the steady-state Kalman filter to the case in which the underlying system is also uncertain.

The class of uncertain systems considered is described by the state equations:
\begin{equation}\label{eq:met_unc_sys_grtd_cost}
\begin{split}
\dot{x}(t) &= [A+D_1\Delta(t)E_1]x(t)+w_1(t), \, x(t_0) = x_0;\\
y(t) &= [C+D_2\Delta(t)E_1]x(t)+w_2(t),
\end{split}
\end{equation}
where $x(t) \in \mathbb{R}^n$ is the state, $y(t) \in \mathbb{R}^l$ is the measured output, $x_0$ is the initial condition which is assumed to be a zero mean Gaussian random vector, $\Delta(t)$ is a time-varying matrix of uncertain parameters satisfying the bound $\Delta^T(t)\Delta(t) \leq I$, and $w_1(t)$ and $w_2(t)$ are zero mean white Gaussian noise processes with joint covariance matrix
\begin{equation}
\left\langle\left[\begin{array}{c}
w_1\\
w_2
\end{array}\right]\left[\begin{array}{cc}
w_1^T & w_2^T
\end{array}\right]\right\rangle = \left[\begin{array}{cc}
V_1 & 0\\
0 & V_2
\end{array}\right] > 0.
\end{equation}

Since we are dealing with the steady-state filtering problem, we assume that the initial time $t_0 \to -\infty$.

\begin{theorem}\label{thm:guar_cost_filter}\cite{PM}
Consider the uncertain system (\ref{eq:met_unc_sys_grtd_cost}), that is assumed to be quadratically stable. Then, there exists a constant $\epsilon^{*}>0$, such that for all $\epsilon\in(0,\epsilon^{*})$, the Riccati equation
\begin{equation}\label{eq:met_riccati1_grtd_cost}
AS+SA^T+\epsilon SE_1^TE_1S+\frac{1}{\epsilon}D_1D_1^T+V_1=0
\end{equation}
has a stabilising solution $S^{+}>0$. For any such $\epsilon$, the Riccati equation
\begin{equation}\label{eq:met_riccati2_grtd_cost}
\begin{split}
(A&-D_1D_2^T(\epsilon V_2+D_2D_2^T)^{-1}C)Q\\
&+Q(A-D_1D_2^T(\epsilon V_2+D_2D_2^T)^{-1}C)^T\\
&+\epsilon QE_1^TE_1Q-\epsilon QC^T(\epsilon V_2+D_2D_2^T)^{-1}CQ\\
&+\frac{1}{\epsilon}D_1(I-D_2^T(\epsilon V_2+D_2D_2^T)^{-1}D_2)D_1^T\\
&+V_1=0
\end{split}
\end{equation}
has a stabilising solution $Q^{+}>0$, such that $Q^{+}\leq S^{+}$. Also, the state estimator
\begin{equation}\label{eq:met_est_grtd_cost}
\begin{split}
\dot{\hat{x}}(t)&=(A+\epsilon Q^{+}E_1^TE_1)\hat{x}(t)+(\epsilon Q^{+}C^T+D_1D_2^T)\\
&\quad\times(\epsilon V_2+D_2D_2^T)^{-1}(y(t)-C\hat{x}(t))
\end{split}
\end{equation}
has the following property: Given any $\delta>0$, there exists a matrix $\tilde{Q}>0$ such that $Q^{+}\leq\tilde{Q}<Q^{+}+\delta I$ and (\ref{eq:met_est_grtd_cost}) is a quadratic guaranteed cost state estimator for the system (\ref{eq:met_unc_sys_grtd_cost}) with cost matrix $\tilde{Q}$.

Conversely, given any quadratic guaranteed cost state estimator for the system (\ref{eq:met_unc_sys_grtd_cost}) with cost matrix $\tilde{Q}$, there exists a constant $\epsilon>0$, such that Riccati equations (\ref{eq:met_riccati1_grtd_cost}) and (\ref{eq:met_riccati2_grtd_cost}) have stabilising solutions $S^{+}>0$ and $Q^{+}>0$ respectively and $Q^{+}<\tilde{Q}$.
\end{theorem}

Moreover, the quadratic guaranteed cost state estimator defined by (\ref{eq:met_riccati2_grtd_cost}) and (\ref{eq:met_est_grtd_cost}) has the following property: The steady-state error covariance matrix at time $t$ satisfies the bound $Q_{\Delta}(t)\leq Q^{+}$ for all admissible uncertainties $\Delta(t)$ \cite{PM}.

Furthermore, the optimal guaranteed cost state estimator for the uncertain system (\ref{eq:met_unc_sys_grtd_cost}) can be obtained by choosing $\epsilon>0$ to minimize Tr$(Q^{+})$, where $Q^{+}>0$ is the stabilising solution to Riccati equation (\ref{eq:met_riccati2_grtd_cost}). This minimization is also subject to the constraint that (\ref{eq:met_riccati1_grtd_cost}) has a stabilising solution $S^{+}>0$ \cite{PM}.

\section{Standard Quantum Limit}\label{sec:ou_sql}
The standard quantum limit (SQL) is set by the minimum error in phase estimation that can be obtained using a perfect heterodyne technique, or in other words, a non-adaptive filtering scheme \cite{RPH1,RPH3,RPH4}. We deduce the minimum error covariance for the case of OU noise process from (\ref{eq:unc_sys_model}) using the standard Kalman filtering approach \cite{RPH4}. The underlying principle used in the analysis is that the heterodyne measurement scheme is equivalent to, and incurs the same noise penalty as the \emph{dual-homodyne} scheme \cite{TW}.

The process and measurement equations for the steady-state Kalman filter, yielding the SQL, are \cite{RPH4}:
\begin{equation}\label{eq:sql_model}
\begin{split}
\mathsf{Process \, Model:} \, \dot{\phi} &= -(\lambda+\mu\delta\lambda)\phi + \sqrt{\kappa}v,\\
\mathsf{Measurement \, Model:} \, \vartheta &= \phi + \frac{1}{2|\alpha|}n_1 + \frac{1}{2|\alpha|}n_2,
\end{split}
\end{equation}
where $n_1$ is the measurement noise of one of the homodyne detectors HD1 of the dual-homodyne measurement, $n_2$ is the noise arising from the vacuum entering the empty port of the input beamsplitter corresponding to the arm having HD1 (refer to Fig.~C1 from Ref.~\cite{RPH1}), and $\vartheta$ determines the net measurement current obtained from the dual-homodyne scheme.

Then, the algebraic Riccati equation required to be solved is:
\begin{equation}
-2(\lambda+\mu\delta\lambda) P_{SQL} - 2|\alpha|^2 P_{SQL}^2 + \kappa = 0,
\end{equation}
where $P_{SQL}$ is the desired error covariance (SQL).

The stabilising solution of the above equation is:
\begin{equation}\label{eq:ou_sql}
P_{SQL} = \frac{-(\lambda+\mu\delta\lambda) + \sqrt{(\lambda+\mu\delta\lambda)^2 + 2\kappa|\alpha|^2}}{2|\alpha|^2}.
\end{equation}

This determines the desired SQL, that we have included in our plots in Section \ref{sec:ou_comp_errors}. In addition, note that the transfer function of the Kalman filter here is:
\begin{equation}\label{eq:ou_sql_tf}
G_{SQL}(s) := \frac{K_{SQL}}{s+\lambda_u+K_{SQL}},
\end{equation}
where $\lambda_u=\lambda+\mu\delta\lambda$ and $K_{SQL}$ is the Kalman gain, found to be $K_{SQL}=-\lambda_u+\sqrt{\lambda_u^2+2\kappa|\alpha|^2}$.

\section{Proof for Robust Filter vs.~SQL}\label{sec:proof_rob_sql}
We start with the obvious inequality:
\small
\begin{align}
&\kappa^2|\alpha|^4\geq 0\\
\begin{split}
\Rightarrow\quad &\kappa^2|\alpha|^4+2\kappa|\alpha|^2\lambda^2(1-\mu)^2+\lambda^4(1-\mu)^4\\
&\quad\geq 2\kappa|\alpha|^2\lambda^2(1-\mu)^2+\lambda^4(1-\mu)^4
\end{split}\\
\begin{split}
\Rightarrow\quad &\left(\kappa|\alpha|^2+\lambda^2(1-\mu)^2\right)^2\\
&\quad\geq \lambda^2(1-\mu)^2\left(\lambda^2(1-\mu)^2+2\kappa|\alpha|^2\right)
\end{split}\\
\begin{split}
\Rightarrow\quad &\kappa|\alpha|^2+\lambda^2(1-\mu)^2\\
&\quad\geq\lambda(1-\mu)\sqrt{\lambda^2(1-\mu)^2+2\kappa|\alpha|^2}
\end{split}\\
\begin{split}
\Rightarrow\quad &2\kappa|\alpha|^2+\lambda^2(1-\mu)^2\\
&\quad-\lambda(1-\mu)\sqrt{\lambda^2(1-\mu)^2+2\kappa|\alpha|^2}\geq \kappa|\alpha|^2
\end{split}\\
\begin{split}
\Rightarrow\quad &4\left(2\kappa|\alpha|^2+\lambda^2(1-\mu)^2\right)+\lambda^2(1-\mu)^2\\
&\quad-4\lambda(1-\mu)\sqrt{\lambda^2(1-\mu)^2+2\kappa|\alpha|^2}\\
&\quad\geq 4\kappa|\alpha|^2+\lambda^2(1-\mu)^2
\end{split}\\
\begin{split}
\Rightarrow\quad &\left(2\sqrt{\lambda^2(1-\mu)^2+2\kappa|\alpha|^2}-\lambda(1-\mu)\right)^2\\
&\quad\geq 4\kappa|\alpha|^2+\lambda^2(1-\mu)^2
\end{split}\\
\begin{split}
\Rightarrow\quad &2\sqrt{\lambda^2(1-\mu)^2+2\kappa|\alpha|^2}-\lambda(1-\mu)\\
&\quad\geq\sqrt{\lambda^2(1-\mu)^2+4\kappa|\alpha|^2}.
\end{split}
\end{align}
\normalsize

\section{Two-time Error Correlations}\label{sec:error_2t}
Here, we devise methods to compute the two-time error correlations of an optimal Kalman filter and also an arbitrary (suboptimal) filter for a given system.

\subsection{Kalman filter}\label{sec:app_opt_filter_2t}
Given the process and measurement models as follows:
\small
\begin{equation}
\begin{split}
\mathsf{Process \, Model:} \, \dot{x}(t) &= Ax(t) + Bv(t), \, x(0) = x_0,\\
\mathsf{Measurement \, Model:} \, y(t) &= Cx(t) + Dw(t),
\end{split}
\end{equation}\normalsize
where
\begin{equation}
\begin{split}
\langle v(t)v^T(\tau)\rangle &= Q\delta(t-\tau),\\
\langle w(t)w^T(\tau)\rangle &= R\delta(t-\tau),\\
\langle v(t)w^T(\tau)\rangle &= 0,\\
\langle v(t)x^T(0)\rangle &= \langle w(t)x^T(0)\rangle = 0,
\end{split}
\end{equation}
the optimal steady-state Kalman filter is given by:
\begin{equation}
\dot{\hat{x}}(t) = A\hat{x}(t) + K[y(t)-C\hat{x}(t)], \qquad \hat{x}(0) = \hat{x}_0.
\end{equation}

Here, $K$ is the Kalman gain, $K = PC^T(DRD^T)^{-1}$, where $P$ is the error covariance (mean-square error), obtained by solving the algebraic Riccati equation:
\begin{equation}\label{eq:kalman_are}
AP+PA^T-PC^T(DRD^T)^{-1}CP+BQB^T = 0.
\end{equation}

Then, the estimation error is
\begin{equation}
\varepsilon(t) = x(t) - \hat{x}(t).
\end{equation}
Subtracting the filter equation from the process model,
\small
\begin{equation}\label{eq:tte_1}
\begin{split}
\dot{x}(t) - \dot{\hat{x}}(t) &= (A-KC)[x(t)-\hat{x}(t)]\\
&\quad+Bv(t)-KDw(t),\\
\mathsf{or, \, \, }\dot{\varepsilon}(t) &= (A-KC)\varepsilon(t)+Bv(t)-KDw(t).
\end{split}
\end{equation}\normalsize

If $\Phi(t,0)$ is the state transition matrix for $(A-KC)$, then the solution of the above equation is
\small
\begin{equation}\label{eq:tte_2}
\begin{split}
\varepsilon(t) = \Phi(t,0)\varepsilon(0)&+\int_0^t\Phi(t,s)Bv(s)ds\\
&-\int_0^t\Phi(t,s)KDw(s)ds.
\end{split}
\end{equation}\normalsize
Post-multiplying (\ref{eq:tte_1}) by $\varepsilon^T(t-\tau)$, where $\tau \geq 0$ and taking the expectation, we get:

\vspace*{-2mm}\small
\begin{equation}\label{eq:tte_3}
\begin{split}
\langle \dot{\varepsilon}(t)\varepsilon^T(t-\tau)\rangle &= (A-KC)\langle \varepsilon(t)\varepsilon^T(t-\tau)\rangle \\
&\quad+ B\langle v(t)\varepsilon^T(t-\tau)\rangle \\
&\quad- KD\langle w(t)\varepsilon^T(t-\tau)\rangle .
\end{split}
\end{equation}\normalsize
Similarly, we obtain:
\small
\begin{equation}\label{eq:tte_4}
\begin{split}
\langle \varepsilon(t)\dot{\varepsilon}^T(t-\tau)\rangle &= \langle \varepsilon(t)\varepsilon^T(t-\tau)\rangle(A-KC)^T\\
&\quad+ \langle \varepsilon(t)v^T(t-\tau)\rangle B^T\\
&\quad- \langle \varepsilon(t)w^T(t-\tau)\rangle D^TK^T.
\end{split}
\end{equation}\normalsize

Define $P(t,t-\tau) := \langle \varepsilon(t)\varepsilon^T(t-\tau)\rangle$. Then, upon adding (\ref{eq:tte_3}) and (\ref{eq:tte_4}), we get:

\small
\begin{equation}\label{eq:tte_5}
\begin{split}
\dot{P}(t,t-\tau) &= (A-KC)P(t,t-\tau)\\
&\quad+ P(t,t-\tau)(A-KC)^T\\
&\quad+ B\langle v(t)\varepsilon^T(t-\tau)\rangle \\
&\quad+ \langle \varepsilon(t)v^T(t-\tau)\rangle B^T\\
&\quad- KD\langle w(t)\varepsilon^T(t-\tau)\rangle \\
&\quad- \langle \varepsilon(t)w^T(t-\tau)\rangle D^TK^T.
\end{split}
\end{equation}\normalsize
From (\ref{eq:tte_2}), we get:
\small
\begin{equation}
\begin{split}
\langle \varepsilon(t)v^T(t-\tau)\rangle &= \int_0^t \Phi(t,s)B\langle v(s)v^T(t-\tau)\rangle ds\\
&= \int_0^t\Phi(t,s)BQ\delta(s-t+\tau)ds\\
&= \begin{cases}
0, & \tau > t > 0,\\
\frac{1}{2}BQ, & \tau \in \{0,t\},\\
\Phi(t,t-\tau)BQ, & t > \tau > 0,
\end{cases}
\end{split}
\end{equation}\normalsize
using the following three properties of the Dirac delta function:
\small
\begin{equation}
\begin{split}
\int_0^q \delta(p) dp &= \int_{-q}^0 \delta(p)dp = \frac{1}{2},\\
\int_{-q}^q \delta(p) dp &= 1,\\
\int_{-q}^q f(p)\delta(p) dp &= f(0),
\end{split}
\end{equation}\normalsize
where $q>0$ and $f(\cdot)$ is some function. Similarly, we get:
\small
\begin{align}
\langle v(t)\varepsilon^T(t-\tau)\rangle &= \begin{cases}
QB^T\Phi^T(t-\tau,t), & \tau > t > 0,\\
\frac{1}{2}QB^T, & \tau \in \{0,t\},\\
0, & t > \tau > 0,
\end{cases}\\
\langle \varepsilon(t)w^T(t-\tau)\rangle &= \begin{cases}
0, & \tau > t > 0,\\
-\frac{1}{2}KDR, & \tau \in \{0,t\},\\
-\Phi(t,t-\tau)KDR, & t > \tau > 0.
\end{cases}\\
\langle w(t)\varepsilon^T(t-\tau)\rangle &= \begin{cases}
-RD^TK^T &\\
\times\Phi^T(t-\tau,t),&\tau > t > 0,\\
-\frac{1}{2}RD^TK^T,&\tau \in \{0,t\},\\
0, & t > \tau > 0.
\end{cases}
\end{align}
\normalsize

\begin{widetext}
Then, (\ref{eq:tte_5}) can be expressed as:
\small
\begin{equation}\label{eq:tte_6}
\dot{P}(t,t-\tau) = \begin{cases}
(A-KC)P(t,t-\tau)+P(t,t-\tau)(A-KC)^T+\Phi(t,t-\tau)BQB^T &\\
+\Phi(t,t-\tau)KDRD^TK^T, & t > \tau > 0,\\
(A-KC)P(t,t-\tau)+P(t,t-\tau)(A-KC)^T+BQB^T+KDRD^TK^T, & \tau \in \{0,t\},\\
(A-KC)P(t,t-\tau)+P(t,t-\tau)(A-KC)^T+BQB^T\Phi^T(t-\tau,t) &\\
+KDRD^TK^T\Phi^T(t-\tau,t), & \tau > t > 0.
\end{cases}
\end{equation}
\normalsize
\end{widetext}

It can be verified that the above equation for $\tau = 0$ reduces to the standard Kalman filter matrix differential Riccati equation.

At steady state, the left-hand side above is zero, yielding (\ref{eq:kalman_are}) at $\tau=0$. Also, the state transition matrix may be computed as follows:
\begin{equation}
\Phi(t,t-\tau) = e^{(A-KC)\tau} = \Phi^{-1}(t-\tau,t).
\end{equation}

Usually, by the notation $P(t,t-\tau)$, we would imply $t \geq \tau \geq 0$, hence the third case in (\ref{eq:tte_6}) will be ignored since the state transition matrix in this case represents backward-time state transition. However, given $t \geq \tau \geq 0$, the third case allows for verifying that $P(t-\tau,t) = P^T(t,t-\tau)$, as expected. Then, the steady-state matrix equation to be solved to obtain $P(t,t-\tau)$ is:

\vspace*{-2mm}\small
\begin{equation}\label{eq:opt_lyap_2t}
\begin{split}
0 &= (A-KC)P(t,t-\tau)+ P(t,t-\tau)(A-KC)^T\\
&\quad+e^{(A-KC)\tau}BQB^T+e^{(A-KC)\tau}KDRD^TK^T.
\end{split}
\end{equation}
\normalsize

\subsection{Arbitrary Filter}\label{sec:app_sub_filter_2t}
Here, we extend Section \ref{sec:lyapunov} to deduce the form of Lyapunov equation required to be solved to obtain the two-time state covariance matrix and therefore the two-time error correlations of an arbitrary filter for a given system state-space model.

Let the system process and measurement models and an arbitrary filter for the system be determined by the following differential equations:
\begin{equation}
\begin{split}
\mathsf{Process \, Model:} \, \dot{x}(t) &= Ax(t) + Bv(t),\\
x(0) &= x_0,\\
\mathsf{Measurement \, Model:} \, y(t) &= Cx(t) + Dw(t),\\
\mathsf{Filter \, Equation:} \, \dot{\hat{x}}(t) &= A_e\hat{x}(t) + B_ey(t)\\
&= A_e\hat{x}(t) + B_eCx(t)\\
&\quad+ B_eDw(t),
\end{split}
\end{equation}
where again
\begin{equation}
\begin{split}
\langle v(t)v^T(\tau)\rangle &= Q\delta(t-\tau),\\
\langle w(t)w^T(\tau)\rangle &= R\delta(t-\tau),\\
\langle v(t)w^T(\tau)\rangle &= 0,\\
\langle v(t)x^T(0)\rangle &= \langle w(t)x^T(0)\rangle = 0,
\end{split}
\end{equation}

We obtain an augmented system as follows from the process model and the filter equation:
\begin{equation}
\dot{\overline{x}}(t) = \overline{A}\overline{x}(t) + \overline{B}\overline{w}(t),
\end{equation}
where
\begin{equation}
\begin{split}
\overline{x}(t) &= \left[\begin{array}{c}
x(t)\\
\hat{x}(t)
\end{array}\right], \qquad \overline{w}(t) = \left[\begin{array}{c}
v(t)\\
w(t)
\end{array}\right],\\
\overline{A} &= \left[\begin{array}{cc}
A & 0\\
B_eC & A_e
\end{array}\right], \qquad \overline{B} = \left[\begin{array}{cc}
B & 0\\
0 & B_eD
\end{array}\right].
\end{split}
\end{equation}

Then, we get for $\tau \geq 0$,
\begin{equation}
\begin{split}
\langle \dot{\overline{x}}(t)\overline{x}^T(t-\tau)\rangle &= \overline{A}\langle \overline{x}(t)\overline{x}^T(t-\tau)\rangle \\
&\quad+ \overline{B}\langle \overline{w}(t)\overline{x}^T(t-\tau)\rangle.
\end{split}
\end{equation}
Similarly,
\begin{equation}
\begin{split}
\langle \overline{x}(t)\dot{\overline{x}}^T(t-\tau)\rangle &= \langle \overline{x}(t)\overline{x}^T(t-\tau)\rangle \overline{A}^T\\
&\quad+ \langle \overline{x}(t)\overline{w}^T(t-\tau)\rangle \overline{B}^T.
\end{split}
\end{equation}

\vspace*{1mm}
The steady-state state covariance matrix of the augmented system is given by $P_S(t,t-\tau) = \langle \overline{x}(t)\overline{x}^T(t-\tau)\rangle$. Then,
\begin{equation}
\begin{split}
\dot{P}_S(t,t-\tau) &= \langle \dot{\overline{x}}(t)\overline{x}^T(t-\tau)\rangle + \langle \overline{x}(t)\dot{\overline{x}}^T(t-\tau)\rangle \\
&= \overline{A}P_S(t,t-\tau) + P_S(t,t-\tau)\overline{A}^T\\
&\quad+ \overline{B}\langle \overline{w}(t)\overline{x}^T(t-\tau)\rangle \\
&\quad+ \langle \overline{x}(t)\overline{w}^T(t-\tau)\rangle \overline{B}^T.
\end{split}
\end{equation}

Note
\begin{equation}
\begin{split}
\langle \overline{w}(t)\overline{w}^T(t-\tau)\rangle &= \left[\begin{array}{cc}
Q & 0\\
0 & R
\end{array}\right]\delta(s-t+\tau)\\
&=: \overline{Q}\delta(s-t+\tau).
\end{split}
\end{equation}

Now, let $\overline{\Phi}(t,0)$ be the state transition matrix for $\overline{A}$. Then, we have:
\begin{equation}
\overline{x}(t) = \overline{\Phi}(t,0)\overline{x}(0) + \int_0^t \overline{\Phi}(t,s)\overline{B}\overline{w}(s)ds.
\end{equation}
Then,
\small
\begin{equation}
\begin{split}
\langle \overline{x}(t)\overline{w}^T(t-\tau)\rangle &= \int_0^t \overline{\Phi}(t,s)\overline{B}\langle \overline{w}(s)\overline{w}^T(t-\tau)\rangle ds\\
&= \int_0^t \overline{\Phi}(t,s)\overline{B}\overline{Q}\delta(s-t+\tau)ds\\
&= \begin{cases}
0, & \tau > t > 0,\\
\frac{1}{2}\overline{B}\overline{Q}, & \tau \in \{0,t\},\\
\overline{\Phi}(t,t-\tau)\overline{B}\overline{Q}, & t > \tau > 0.
\end{cases}
\end{split}
\end{equation}
\normalsize
Similarly,
\small
\begin{equation}
\langle \overline{w}(t)\overline{x}^T(t-\tau)\rangle = \begin{cases}
0, & t > \tau > 0,\\
\frac{1}{2}\overline{Q}\overline{B}^T, & \tau \in \{0,t\},\\
\overline{Q}\overline{B}^T\overline{\Phi}^T(t-\tau,t), & \tau > t > 0.
\end{cases}
\end{equation}
\normalsize

\begin{widetext}
Thus, we get:
\begin{equation}
\dot{P}_S(t,t-\tau) = \begin{cases}
\overline{A}P_S(t,t-\tau)+P_S(t,t-\tau)\overline{A}^T+\overline{B}\overline{Q}\overline{B}^T\overline{\Phi}^T(t-\tau,t), & \tau > t > 0,\\
\overline{A}P_S(t,t-\tau)+P_S(t,t-\tau)\overline{A}^T+\overline{B}\overline{Q}\overline{B}^T, & \tau \in \{0,t\},\\
\overline{A}P_S(t,t-\tau)+P_S(t,t-\tau)\overline{A}^T+\overline{\Phi}(t,t-\tau)\overline{B}\overline{Q}\overline{B}^T, & t > \tau > 0,
\end{cases}
\end{equation}
where
\begin{equation}
\overline{\Phi}(t,t-\tau) = e^{\overline{A}\tau} = \overline{\Phi}^{-1}(t-\tau,t).
\end{equation}
\end{widetext}

As before, by the notation $P_S(t,t-\tau)$, we would imply $t \geq \tau \geq 0 $, and so, the first case above will be ignored. Then, at steady state, $\dot{P}_S(t,t-\tau) = 0$, and the form of Lyapunov equation required to be solved to obtain the desired two-time state covariance matrix is:
\begin{equation}\label{eq:sub_lyap_2t}
\overline{A}P_S(t,t-\tau)+P_S(t,t-\tau)\overline{A}^T+e^{\overline{A}\tau}\overline{B}\overline{Q}\overline{B}^T = 0.
\end{equation}
Then, noting that the filter estimation error is $\varepsilon(t) = \left[\begin{array}{cc} 1 & -1 \end{array}\right]\overline{x}(t)$, the two-time error correlation for the filter is obtained as follows:
\begin{equation}\label{eq:sub_mse_2t}
\begin{split}
P(t,t-\tau) &= \langle \varepsilon(t)\varepsilon^T(t-\tau)\rangle \\
&= \left[\begin{array}{cc} 1 & -1 \end{array}\right] P_S(t,t-\tau) \left[\begin{array}{c} 1\\ -1 \end{array}\right].
\end{split}
\end{equation}

\section{Matching Two-Time Correlations}\label{sec:more_analysis}
Here, we strive to arrive at a condition under which the two-time error correlations of a suboptimal filter for the uncertain OU phase (\ref{eq:unc_sys_model}) possibly match up with those of the optimal filter for the uncertain OU phase with added extra OU noise (\ref{eq:add_ou_model}).

With reference to Appendix \ref{sec:app_sub_filter_2t}, in our case we have:
\begin{equation}
\overline{A} = \left[\begin{array}{cc}
-\lambda_u & 0\\
B_e & A_e
\end{array}\right], \quad \overline{B} = \left[\begin{array}{cc}
\sqrt{\kappa} & 0\\
0 & \frac{B_e}{2|\alpha|}
\end{array}\right], \quad \overline{Q} = I,
\end{equation}
where $A_e = -(\lambda + K)$ and $B_e = K$ for the Kalman filter from (\ref{eq:kalman_filter}), and $A_e = -\lambda-4|\alpha|^2Q^{+}+\epsilon_{\rm opt}\lambda^2Q^{+}$ and $B_e = 4|\alpha|^2Q^{+}$ for the robust filter from (\ref{eq:robust_filter}). Then, we have:
\begin{equation}
e^{\overline{A}\tau} = \left[\begin{array}{cc}
e^{-\lambda_u\tau} & 0\\
\frac{B_e(e^{A_e\tau}-e^{-\lambda_u\tau})}{A_e+\lambda_u} & e^{A_e\tau}
\end{array}\right].
\end{equation}

Let us denote
\begin{equation}
P_S := \left[\begin{array}{cc}
P_{S1} & P_{S2}\\
P_{S3} & P_{S4}
\end{array}\right].
\end{equation}
Then, we get from (\ref{eq:sub_lyap_2t}) the following:
\small
\begin{align}
-2\lambda_u P_{S1} + e^{-\lambda_u\tau}\kappa &= 0,\\
(-\lambda_u+A_e)P_{S2} + B_eP_{S1} &= 0,\\
(-\lambda_u+A_e)P_{S3} + B_eP_{S1} + \frac{B_e\kappa(e^{A_e\tau}-e^{-\lambda_u\tau})}{A_e+\lambda_u} &= 0,\\
B_e(P_{S2}+P_{S3}) + 2A_eP_{S4} + e^{A_e\tau}\frac{B_e^2}{4|\alpha|^2} &= 0.
\end{align}
\normalsize
Upon solving the above equations, we get:
\begin{align}
P_{S1} &= \frac{e^{-\lambda_u\tau}\kappa}{2\lambda_u},\\
P_{S2} &= -\frac{e^{-\lambda_u\tau}\kappa B_e}{2\lambda_u(A_e-\lambda_u)},\\
P_{S3} &= -\frac{B_e\kappa e^{-\lambda_u\tau}(A_e-\lambda_u)+2B_e\kappa\lambda_ue^{A_e\tau}}{2\lambda_u(A_e^2-\lambda_u^2)},\\
\begin{split}
P_{S4} &= \frac{\kappa B_e^2e^{-\lambda_u\tau}}{2\lambda_u(A_e^2-\lambda_u^2)}\\
&\quad+\frac{B_e^2e^{A_e\tau}(4|\alpha|^2\kappa+\lambda_u^2-A_e^2)}{8|\alpha|^2A_e(A_e^2-\lambda_u^2)}.
\end{split}
\end{align}
Then, from (\ref{eq:sub_mse_2t}), we get $P_\tau := P(t,t-\tau)$:
\begin{equation}\label{eq:add_ou_sub_corr_2t}
\begin{split}
P_\tau &= P_{S1}-P_{S2}-P_{S3}+P_{S4}\\
&= e^{-\lambda_u\tau}\frac{\kappa\left((A_e+B_e)^2-\lambda_u^2\right)}{2\lambda_u(A_e^2-\lambda_u^2)}\\
&\quad-e^{A_e\tau}\frac{B_e^2}{8|\alpha|^2A_e}+e^{A_e\tau}\frac{B_e\kappa(2A_e+B_e)}{2A_e(A_e^2-\lambda_u^2)}.
\end{split}
\end{equation}

Next, with reference to Appendix (\ref{sec:app_opt_filter_2t}), we rewrite (\ref{eq:opt_lyap_2t}) with tildes to avoid ambiguity with notation here:
\small
\begin{equation}\label{eq:opt_lyap_tilde_2t}
\begin{split}
0 &= (A-KC)\tilde{P}_\tau+\tilde{P}_\tau(A-KC)^T\\
&\quad+e^{(A-KC)\tau}BQB^T+e^{(A-KC)\tau}KDRD^TK^T.
\end{split}
\end{equation}
\normalsize
Here, the matrices are as defined in (\ref{eq:add_ou_matrices}), $K = PC^T(DRD^T)^{-1}$, where $P$ is as obtained from (\ref{eq:add_ou_riccati}), and $Q=I$ and $R=1$. Note that $P = \tilde{P}_{\tau=0} = \tilde{P}(t,t)$. For simplicity, let $K := \left[\begin{array}{cc}K_1 & K_2\end{array}\right]^T$. Clearly, from (\ref{eq:add_ou_err_cov}), we have:
\begin{equation}\label{eq:add_ou_2t_k12}
\begin{split}
K_1 &= 4|\alpha|^2(P_1+P_2) = \frac{\kappa(-\lambda_u+\beta)}{\kappa+\kappa_n},\\
K_2 &= 4|\alpha|^2(P_2+P_3) = \frac{\kappa_n(-\lambda_u+\beta)}{\kappa+\kappa_n}.\\
\end{split}
\end{equation}
Then,
\begin{equation}
A-KC = \left[\begin{array}{cc}
-\lambda_u-K_1 & -K_1\\
-K_2 & -\lambda_u-K_2
\end{array}\right],
\end{equation}
and the matrix exponential is obtained as:
\scriptsize
\begin{equation}
e^{(A-KC)\tau} = \left[\begin{array}{cc}
\frac{e^{-A_n\tau}K_1+e^{-\lambda_u\tau}K_2}{K_1+K_2} & \frac{(e^{-A_n\tau}-e^{-\lambda_u\tau})K_1}{K_1+K_2}\\
\frac{(e^{-A_n\tau}-e^{-\lambda_u\tau})K_2}{K_1+K_2} & \frac{e^{-\lambda_u\tau}K_1+e^{-A_n\tau}K_2}{K_1+K_2}
\end{array}\right],
\end{equation}
\normalsize
where $A_n=\lambda_u+K_1+K_2$.

Let
\begin{equation}
\tilde{P}_\tau := \left[\begin{array}{cc}
\tilde{P}_1 & \tilde{P}_2\\
\tilde{P}_3 & \tilde{P}_4
\end{array}\right].
\end{equation}

We then wish to find an added noise such that the two-time error correlations described by $\tilde{P}_1$ from above equation match those for the suboptimal filter given by $P_\tau$ from (\ref{eq:add_ou_sub_corr_2t}). The only parameter available to vary for the noise is $\kappa_n$, which is independent of $\tau$.

Expanding (\ref{eq:opt_lyap_tilde_2t}) and solving for $\tilde{P}_1$, the desired two-time error correlation for the optimal filter, we get:
\begin{equation}\label{eq:add_ou_opt_corr_2t}
\tilde{P}_1 = e^{-\lambda_u\tau}\frac{\kappa\kappa_n}{2\lambda_u(\kappa+\kappa_n)}+e^{-\beta\tau}\frac{\kappa^2(-\lambda_u+\beta)}{4|\alpha|^2(\kappa+\kappa_n)^2}.
\end{equation}
Comparing (\ref{eq:add_ou_sub_corr_2t}) and (\ref{eq:add_ou_opt_corr_2t}), we see that we need the following to hold:
\small
\begin{align}
A_e &= -\beta = -\sqrt{4|\alpha|^2(\kappa+\kappa_n)+\lambda_u^2},\label{eq:add_ou_2t_cond1}\\
&\frac{(A_e+B_e)^2-\lambda_u^2}{A_e^2-\lambda_u^2}=\frac{\kappa_n}{\kappa+\kappa_n},\label{eq:add_ou_2t_cond2}\\
&\frac{B_e\kappa(2A_e+B_e)}{2A_e(A_e^2-\lambda_u^2)}-\frac{B_e^2}{8|\alpha|^2A_e}=\frac{\kappa^2(-\lambda_u+\beta)}{4|\alpha|^2(\kappa+\kappa_n)^2}.\label{eq:add_ou_2t_cond3}
\end{align}
\normalsize
It turns out that it is not possible to choose a suitable value for only $\kappa_n$ that will allow for all the three independent conditions above (i.e.~(\ref{eq:add_ou_2t_cond1}), (\ref{eq:add_ou_2t_cond2}) and (\ref{eq:add_ou_2t_cond3})) to be satisfied simultaneously. Further, if we consider there is another parameter $\lambda_n$ available to vary for the added noise, in addition to $\kappa_n$, (\ref{eq:add_ou_opt_corr_2t}) will have three (instead of two) time constants, required to be matched with two time constants from (\ref{eq:add_ou_sub_corr_2t}). Thus, it does not seem possible to match up the two-time error correlations.

\end{document}